\documentclass[]{spie}  

\usepackage{amsmath,amsfonts,amssymb}
\usepackage{graphicx}
\usepackage[colorlinks=true, allcolors=blue]{hyperref}

\usepackage{lineno}

\title{Development of the Low Frequency Telescope Focal Plane Detector Modules for LiteBIRD}

\author[a,b]{B. Westbrook}
\author[a]{C. Raum}
\author[a]{S. Beckman}
\author[a,b]{A. T. Lee}
\author[a]{N. Farias}
\author[a]{A. Bogdan}
\author[a,c]{A. Hornsby}
\author[c]{A. Suzuki}
\author[c]{K. Rotermund}
\author[c]{T. Elleflot}
\author[d]{J. E. Austermann}
\author[d]{J. A. Beall}
\author[d]{S. M. Duff}
\author[d]{J. Hubmayr}
\author[d]{M. R. Vissers}
\author[d]{M. J. Link}
\author[e]{G. Jaehnig}
\author[e]{N. Halverson}
\author[f]{T. Ghigna}
\author[f,g,h,i]{M. Hazumi}
\author[g]{S. Stever} 
\author[h]{Y. Minami}
\author[i]{K. L. Thompson}
\author[j]{M. Russell}
\author[j]{K. Arnold}
\author[j]{M. Silva-Feaver}

\author[]{for the LiteBIRD collaboration}

\affil[a]{UC Berkeley Physics Department, 151 Physics North, Berkeley, USA}
\affil[b]{Radio Astronomy Lab, UC Berkeley, Berkeley, USA}
\affil[c]{Lawrence Berkeley National Lab, Berkeley, USA}
\affil[d]{National Institute of Standards and Technology, Boulder, USA}
\affil[e]{Colorado University, Boulder, Boulder, USA}
\affil[f]{Kavli IPMU (WPI), UTIAS, The University of Tokyo, Kashiwa, Chiba 277-8583, Japan}
\affil[g]{Okayama University Faculty of Science, Okayama, Japan}
\affil[h]{Research Center for Nuclear Physics (RCNP), Osaka University, Osaka, Japan}
\affil[i]{Stanford University and Kavli Institute for Particle Astrophysics and Cosmology, Stanford, USA}
\affil[j]{UC San Diego Physics Department, Berkeley, USA}

\authorinfo{Further author information: (Send correspondence to Benjamin G. Westbrook)\\E-mail: bwestbrook@berkeley.edu, Telephone: 1 415 308 6193}

\usepackage[amssymb]{SIunits}
\usepackage{geometry}               
\usepackage[parfill]{parskip}       
\usepackage{graphicx}
\usepackage{subfiles}
\usepackage{makecell}
\usepackage[table]{xcolor}
\usepackage{amssymb}
\usepackage{epstopdf}
\usepackage[utf8]{inputenc}
\usepackage[english]{babel}
\usepackage{url}
\usepackage{array}
\usepackage{subfigure}
\usepackage{multirow}
\usepackage{float}
\usepackage{longtable}
\usepackage{mathtools}
\usepackage{mathrsfs}
\usepackage{indentfirst}
\usepackage{pdfpages}
\usepackage{xcolor}
\usepackage{glossaries}
\makeglossaries

\newacronym{fpu}{FPU}{Focal Plane Unit} 
\newacronym{fpm}{FPM}{Focal Plane Module} 
\newacronym{fps}{FPS}{Focal Plane Structure}
\newacronym{fph}{FPH}{Focal Plane Hood}
\newacronym{cru}{CRU}{Cryogenic Readout Units}
\newacronym{crh}{CRH}{Cryogenic Readout Harness} 
\newacronym{cr}{CRE}{Cryogenic Readout Electronics}
\newacronym{wr}{WR}{Warm Readout}

\newacronym{hf}{HF}{High-Frequency}
\newacronym{lf}{LF}{Low-Frequency}
\newacronym{mf}{MF}{Mid-Frequency}

\newacronym{hft}{HFT}{High-Frequency Telescope}
\newacronym{mft}{MFT}{Mid-Frequency Telescope}
\newacronym{lft}{LFT}{Low-Frequency Telescope}
\newacronym{mhft}{MHFT}{Mid- and High-Frequency Telescopes}

\newacronym{lffpu}{LF-FPU}{Low-Frequency Focal Plane Unit}
\newacronym{mffpu}{MF-FPU}{Mid-Frequency Focal Plane Unit}
\newacronym{hffpu}{HF-FPU}{High-Frequency Focal Plane Unit}
\newacronym{mffpm}{MF-FPM}{Mid-Frequency Focal Plane Module}
\newacronym{hffpm}{HF-FPM}{High-Frequency Focal Plane Module}

\newacronym{lffpm}{LF-FPM}{Low-Frequency Focal Plane Module}
\newacronym{lffps}{LF-FPS}{Low-Frequeny Focal Plane Structure}
\newacronym{lffph}{LF-FPH}{Low-Frequency Focal Plane Hood}
\newacronym{lfcrh}{LF-CRH}{Low-Frequency Cryogenic Readout Harness}

\newacronym{2kadrfs}{2K-ADRCC}{2K Adiabatic Demagnetization Refrigerator cryocooler and controller}
\newacronym{2kadr}{2K-ADR}{Adiabatic Demagnetization Refrigerator that cools the 2K stage}
\newacronym{skadr}{Sub-K ADR}{Adiabatic Demagnetization Refrigerator that cools the Sub-Kelvin stages}
\newacronym{dr}{DR}{Dilution Refridgerator}
\newacronym{adr}{ADR}{Adiabatic Demagnetization Refrigerator}
\newacronym{adrc}{ADRC}{Adiabatic Demagnetization Refrigerator Controller}
\newacronym{jt2}{JT2}{Joule-Thomson 2\,K cooled stage}
\newacronym{jt4}{JT5}{Joule-Thomson 5\,K}
\newacronym{jt}{JT}{Joule-Thomson}
\newacronym{em}{EM}{engineering model}
\newacronym{fm}{FM}{flight model}
\newacronym{dm}{DM}{Development Model}

\newacronym{mars}{MARS}{Metamaterial Anti-Reflection Surface}
\newacronym{arc}{ARC}{Anti-Reflection Coating}
\newacronym{crmscb}{BCRML}{Backside Cosmic Ray Mitigation Layer}
\newacronym{ald}{ALD}{Atomic Layer Deposition}
\newacronym{rn}{$R_n$}{Normal Resistance}
\newacronym{psat}{$P_{sat}$}{Saturation Power}
\newacronym{tc}{$T_c$}{Superconducting Transition Temperature}
\newacronym{tb}{$T_b$}{Bath Temperature}
\newacronym{t0}{$\tau_0$}{Intrinsic Time Constant}

\newacronym{fte}{FTE}{ful-time equivalent}
\newacronym{csr}{CSR}{Concept Study Report}
\newacronym{am}{AM}{Assurance Manager}
\newacronym{pdam}{PDAM}{Project Deputy Assurance Manager}
\newacronym{srr}{SRR}{System Requirement Review}
\newacronym{mdr}{MDR}{Mission Definition Review}
\newacronym{pdr}{PDR}{Preliminary Design Review}
\newacronym{cdr}{CDR}{Critical Design Review}
\newacronym{jcdr}{JCDR}{JAXA Concept Design Report}
\newacronym{mel}{MEL}{Master Equipment List}
\newacronym{empfm}{EM/PFM}{Engineering Model / Pre-Flight-Model} 
\newacronym{icd}{ICD}{Interface Control Documents}
\newacronym{sir}{SIR}{System Integration Review}
\newacronym{trlpr}{TRLPR}{Technology Readiness Level Path Review}
\newacronym{plar}{PLAR}{Post-Launch Assessment Review}
\newacronym{pi}{PI}{Principal Investigator}
\newacronym{we}{WE}{Warm Electronics}
\newacronym{wbs}{WBS}{Work Breakdown Structure}
\newacronym{evm}{EVM}{Earned Value Management}
\newacronym{rtm}{RTM}{Requirements and Traceability Matrix}
\newacronym{stm}{STM}{Science Traceability Matrix}
\newacronym{nte}{NTE}{Not-To-Exceed}
\newacronym{mam}{MAM}{Mission Assurance Manager}
\newacronym{dmam}{DMAM}{Deputy Mission Assurance Manager}
\newacronym{cmp}{CMP}{Configuration Management Plan}
\newacronym{jset}{JSET}{Joint Systems Engineering Team}
\newacronym{it}{I\&T}{Integration \& Test}
\newacronym{fmea}{FMEA}{failure modes and effects analysis}
\newacronym{fta}{FTA}{Fault Tree Analysis}
\newacronym{sma}{SMA}{Safety Mission Assurance}
\newacronym{semp}{SEMP}{Systems Engineering Management Plan}
\newacronym{ccb}{CCB}{Configuration Control Board}
\newacronym{srb}{SRB}{Standing Review Board}
\newacronym{tbd}{TBD}{to be done}
\newacronym{tbr}{TBR}{to be resolved}
\newacronym{cbe}{CBE}{current best estimate}
\newacronym{mev}{MEV}{maximum expected value}
\newacronym{plm}{PLM}{Payload Module}
\newacronym{paip}{PAIP}{Performance Assurance Implementation Plan}
\newacronym{orr}{ORR}{Operational Readiness Review}
\newacronym{l1}{L1}{Level 1}
\newacronym{l2}{L2}{Level 2}
\newacronym{l3}{L3}{Level 3}
\newacronym{l4}{L4}{Level 4}
\newacronym{l5}{L5}{Level 5}
\newacronym{pmo}{PMO}{Partner Mission of Opportunity}
\newacronym{mo}{MO}{Mission of Opportunity}
\newacronym{ac}{AC}{Advisory Committee}
\newacronym{trl}{TRL}{Technology Readiness Level}
\newacronym{dcr}{DCR}{Data Completeness Review}
\newacronym{moc}{MOC}{Mission Operations Center}
\newacronym{soc}{SOC}{Science Operations Center}
\newacronym{boe}{BoE}{Basis of Estimate}
\newacronym{mdra}{MDRA}{Mission Definition Requirements Agreement}
\newacronym{poc}{POC}{point of contact}

\newacronym{fts}{FTS}{Fourier Transform Spectrometer}
\newacronym{tod}{TOD}{Time-Ordered Data}
\newacronym{cib}{CIB}{Cosmic Infrared Background}
\newacronym{sts}{STS}{star tracker system}
\newacronym{drie}{DRIE}{deep reactive ion etching}
\newacronym{css}{CSS}{coarse sun sensor}
\newacronym{iru}{IRU}{inertial reference unit}
\newacronym{gse}{GSE}{ground system equipment}
\newacronym{sq}{SQUID}{Superconducting Quantum Interference Device}
\newacronym{scu}{SCU}{SQUID Controller Unit}
\newacronym{dlfov}{DLFOV}{Diffraction-Limited Field of View}
\newacronym{hwp}{HWP}{Half-Wave Plate}
\newacronym{tes}{TES}{Transition-Edge Sensor}
\newacronym{fov}{FOV}{Field of view}
\newacronym{jsg}{JSG}{joint study group}
\newacronym{ncr}{NCR}{noise to carrier ratio}
\newacronym{ar}{AR}{Anti-Reflection}
\newacronym{rwa}{RWA}{reaction wheel assembly}
\newacronym{omt}{OMT}{orthomode transducer}
\newacronym{dac}{DAC}{digital-to-analog converter}
\newacronym{adc}{ADC}{analog-to-digital converter}
\newacronym{cmb}{CMB}{Cosmic Microwave Background}
\newacronym{did}{DID}{data item description}
\newacronym{saa}{SAA}{SQUID Array Amplifier}
\newacronym{dpu}{DPU}{Digital Processing Unit}
\newacronym{pcb}{PCB}{Printed Circuit Board}
\newacronym{co}{CO}{Carbon Monoxide}
\newacronym{rf}{RF}{Radio-Frequency}
\newacronym{lc}{LC}{inductor/capacitor}
\newacronym{dfmux}{DfMux}{digital frequency-domain multiplexing}
\newacronym{net}{NET}{Noise Equivalent Temperature}
\newacronym{fpga}{FPGA}{Field Programmable Gate Array}
\newacronym{fll}{FLL}{Flux Locked Loop}
\newacronym{dan}{DAN}{Digital Active Nulling}
\newacronym{vhdl}{VHDL}{very high speed integrated circuit hardware description language}
\newacronym{asic}{ASIC}{Application Specific Integrated Circuit}
\newacronym{nep}{NEP}{Noise Equivalent Power}
\newacronym{cpw}{CPW}{Coplanar-Waveguide}
\newacronym{cvd}{CVD}{chemical vapor deposition}
\newacronym{pecvd}{PECVD}{plasma enhanced chemical vapor deposition}
\newacronym{lpcvd}{LPCVD}{low pressure chemical vapor deposition}
\newacronym{lsn}{LSN}{low Stress nitride}
\newacronym{rga}{RGA}{Residual Gas Analyzer}
\newacronym{ecr}{ECR}{electron cyclotron resonance}
\newacronym{di}{DI}{de-ionized}
\newacronym{pan}{PAN}{Phosphoric, Acetic, Nitirc}
\newacronym{afm}{AFM}{Atomic Force Microscopy}
\newacronym{sem}{SEM}{Scanning Electron Microscopy}
\newacronym{mems}{MEMS}{MicroElectroMechanical Systems}
\newacronym{ms}{MS}{Microstrip}
\newacronym{mkid}{MKID}{Microwave Kinetic Inductance Detector}
\newacronym{ame}{AME}{Anomalous Microwave Emission}
\newacronym{toast}{TOAST}{Time Ordered Astrophysics Scalable Tools}
\newacronym{cfrp}{CFRP}{carbon fiber reinforced plastic}
\newacronym{cad}{CAD}{Computer Automated Design}
\newacronym{hfss}{HFSS}{High Frequency Simulation Software}
\newacronym{cte}{CTE}{co-efficient of thermal expansion}

\newacronym{carma}{CARMA}{Combined Array for Research in Millimeter-wave Astronomy}
\newacronym{alma}{ALMA}{Atacama Large Millimeter/sub-millimeter Array}
\newacronym{cmbs4}{CMB-S4}{CMB Stage-4}
\newacronym{so}{SO}{Simons Observatory}

\newacronym{sa}{SA}{Simons Array}
\newacronym{sptpol}{SPT-Pol}{South Pole Telescope Polarization Experiment}
\newacronym{spt3g}{SPT-3G}{South Pole Telescope Third Generation}
\newacronym{spo}{SPO}{South Pole Observatory}
\newacronym{cobe}{COBE}{Cosmic Background Explorer}
\newacronym{wmap}{WMAP}{Wilkinson Microwave Anisotropy Probe}
\newacronym{hfi}{HFI}{High Frequency Instrument}
\newacronym{xrism}{XRISM}{X-ray Imaging and Spectroscopy Mission}
\newacronym{planck}{Planck}{Planck}
\newacronym{gpb}{GPB}{Gravity Probe B}
\newacronym{spica}{SPICA}{Space Infrared Telescope for Cosmology and Astrophysics}
\newacronym{cdms}{CDMS}{Cryogenic Dark Matter Search}
\newacronym{safari}{SAFARI}{SpicA FAR-infrared Instrument}
\newacronym{htt}{HTT}{Huan Tran Telescope}

\newacronym{ptep}{PTEP}{Progress of Theoretical and Experimental Physics}
\newacronym{cu}{CU}{University of Colorado, Boulder}
\newacronym{lasp}{CU/LASP}{CU-Boulder Laboratory for Atmospheric and Space Physics}
\newacronym{ucsd}{UC San Diego}{University of California, San Diego}
\newacronym{nist}{NIST}{the National Institute of Standards and Technologies}
\newacronym{uw}{UC San Diego}{the University of California, San Diego}
\newacronym{ucb}{UC Berkeley}{the University of California, Berkeley}
\newacronym{ccc}{$C^3$}{Computational Cosmology Center}
\newacronym{lbnl}{LBNL}{Lawrence Berkeley National Laboratory}
\newacronym{hpc}{HPC}{High Performance Computing}
\newacronym{ssl}{UCB/SSL}{UC Berkeley Space Sciences Laboratory}
\newacronym{cnes}{CNES}{National Centre for Space Studies}
\newacronym{esa}{ESA}{European Space Agency}
\newacronym{great}{GREAT}{GRound station for deep space Exploration And Tele-communication}
\newacronym{csa}{CSA}{Canadian Space Agency}
\newacronym{kek}{KEK}{High Energy Accelerator Research Organization in Tsukuba, Japan}
\newacronym{ut}{U of T}{University of Tokyo}
\newacronym{nersc}{NERSC}{National Energy Research Scientific Computing center}
\newacronym{nasa}{NASA}{National Aeronautics and Space Administration}
\newacronym{nsf}{NSF}{National Science Foundation}
\newacronym{eu}{EU}{Euopean Union}
\newacronym{jaxa}{JAXA}{Japan Aerospace Exploration Agency}
\newacronym{gsfc}{GSFC}{Goddard Space Flight Center}
\newacronym{ipmu}{IPMU}{Kavli Institute for the Physics and Mathematics of the Universe}
\newacronym{nec}{NEC}{Nippon Electric Corporation}
\newacronym{lambda}{LAMBDA}{Legacy Archive for Microwave Background Data Analysis}
\newacronym{shi}{SHI}{Sumitomo Heavy Industries}
\newacronym{stanford}{Stanford}{Stanford University}
\newacronym{cea}{CEA}{French Alternative Energies and Atomic Energy Commission}
\newacronym{apc}{APC}{Laboratoire Astroparticule et Cosmologie}
\newacronym{isas}{ISAS}{Institute of Space and Astronautical Science}
\newacronym{mnl}{MNL}{Marvell Nanofabrication Laboratory}
\newacronym{bmf}{BMF}{Boulder Microfabrication Facility}
\newacronym{sccm}{SCCM}{standard cubic centimeters per minute}

\newacronym{pse}{PSE}{Project System Engineer}
\newacronym{pset}{PSET}{Project System Engineer Team}
\newacronym{pm}{PM}{Project Manager}
\newacronym{pfm}{PFM}{Project Financial Manager}

\newacronym{rpm}{RPM}{revolutions per minute}
\newacronym{rms}{RMS}{root mean square}

\newacronym{ci}{CI}{cold intermediate}
\newacronym{wi}{WI}{warm intermediate}

\def\LB{\textit{LiteBIRD}}
\def\lb{\textit{LiteBIRD}}

\newcolumntype{L}[1]{>{\raggedright\let\newline\\\arraybackslash\hspace{0pt}}m{#1}}
\newcolumntype{C}[1]{>{\centering\let\newline\\\arraybackslash\hspace{0pt}}m{#1}}
\newcolumntype{R}[1]{>{\raggedleft\let\newline\\\arraybackslash\hspace{0pt}}m{#1}}
\newcolumntype{N}{@{}m{0pt}@{}}
\def\pixelred{\color[RGB]{127, 9, 9}}
\def\pixelyellow{\color[RGB]{103, 102, 1}}
\def\pixelgreen{\color[RGB]{1, 102, 1}}
\def\pixelblue{\color[RGB]{3, 4, 120}}

\date{\today}                                           

\begin{document}
\maketitle
\pagestyle{empty}


\begin{abstract}
\lb~is a JAXA-led strategic large-class satellite mission designed to measure the polarization of the cosmic microwave background and Galactic foregrounds from 34 to 448 GHz across the entire sky from L2 in the late 2020s. The scientific payload includes three telescopes which are called the low-, mid-, and high-frequency telescopes each with their own receiver that covers a portion of the mission's frequency range.  The low frequency telescope will map synchrotron radiation from the Galactic foreground and the cosmic microwave background.  We discuss the design, fabrication, and characterization of the low-frequency focal plane modules for low-frequency telescope, which has a total bandwidth ranging from 34 to 161~GHz. There will be a total of 4 different pixel types with 8 overlapping bands to cover the full frequency range.  These modules are housed in a single low-frequency focal plane unit which provides thermal isolation, mechanical support, and radiative baffling for the detectors.  The module design implements multi-chroic lenslet-coupled sinuous antenna arrays coupled to transition edge sensor bolometers read out with frequency-domain mulitplexing.  While this technology has strong heritage in ground-based cosmic microwave background experiments,  the broad frequency coverage, low optical loading conditions, and the high cosmic ray background of the space environment require further development of this technology to be suitable for \LB. In these proceedings, we discuss the optical and bolometeric characterization of a triplexing prototype pixel with bands centered on 78, 100, and 140 GHz.

\end{abstract}

\section{Introduction}

\LB\ is a  JAXA-led strategic Large-Class satellite mission that will map the polarization of the \gls{cmb} and cosmic foreground over the entire sky with an angular resolution appropriate to cover the multipole range $2 \leq \ell \leq 200$.  The payload consists of three telescopes called the \gls{lft}, \gls{mft}, and \gls{hft} each of which has a corresponding \gls{fpu}. It is the primary responsibility of the US team to deliver the \gls{fm} \gls{fpu}s for the mission. \lb\ will deploy a total of 15 detector bands ranging from 34 to 448~GHz which are distributed across the \gls{lffpu}, \gls{mffpu}, and \gls{hffpu}.  An overview of  \lb\ can be found in the journal of the \gls{ptep} \cite{LiteBIRD:2022_PTEP, LiteBIRD:2021_Hazumi}.

 The \gls{lffpu} holds eight square 140~mm x 140~mm tiles with four distinct pixel types and two distinct \gls{fpm} types with a pixel pitch of 32~mm for the lowest frequency pixels and  16~mm for the highest frequency pixels.  The \gls{lffpu} has a frequency coverage of 34 to 161 GHz making it sensitive to both synchrotron radiation and the \gls{cmb}.  These proceedings will discuss the development of the \glspl{fpm} for the \gls{lffpu}, including the optical interfaces to the telescope, the coupling to the \gls{dfmux} readout, implementation of cosmic ray mitigation, and the development of one of the triplexing pixels in the \gls{lft}.

The \gls{lffpu} is being developed at \gls{ucb} \gls{mnl} in close collaboration with researchers at \gls{bmf} at \gls{nist} \cite{Jaehnig:2022_SPIE}. Both institutions have extensive heritage in fabricating \gls{tes} bolometer arrays for studies of the polarization of the \gls{cmb} for both ground-based and balloon-borne experiments making them especially suitable for \lb. The \gls{mnl} is a $\sim$10,000 square foot class~100 clean room that has rich heritage of fabricating detectors for \gls{cmb} experiments including: APEX-SZ, SPT-SZ, EBEX, POLARBEAR-1, \gls{sa}, and \gls{so} \cite{Bleem:2015_SPTSZGalaxies, Abitbol:2018_EBEX, Arnold:2010_POLARBEAR, Suzuki:2016_PB2SA, Abitbol:2019_SODecadal}. The \gls{kek} will provide the \gls{lffpu} to the mission while the US will be in charge of developing and testing the \glspl{fpm} for the mission.

In addition to the work reported in these proceedings, there are other relevant developments for \lb\ reported at this conference.   More information on development of \gls{lft} and \gls{mft} detector  arrays at \gls{nist} and \gls{cu} is in SPIE 2022 proceedings by Jaehnig. \cite{Jaehnig:2022_SPIE}.   One can find information on the susceptibility of \lb\ \glspl{tes} to magnetic fields in SPIE 2022 proceedings by Ghigna\cite{Ghinga:2022_SPIE}.  One can also find more details on \gls{cr} development for \lb\ in SPIE 2022 proceedings by Russell\cite{Russell:2022_SPIE}.

\section{Coupling Design}
\label{sec:couplingDesign}

\begin{figure}[ht!]
\centering
\subfigure[]
{\label{subfig:lf4Fpm}
\includegraphics[width = 0.43\textwidth]{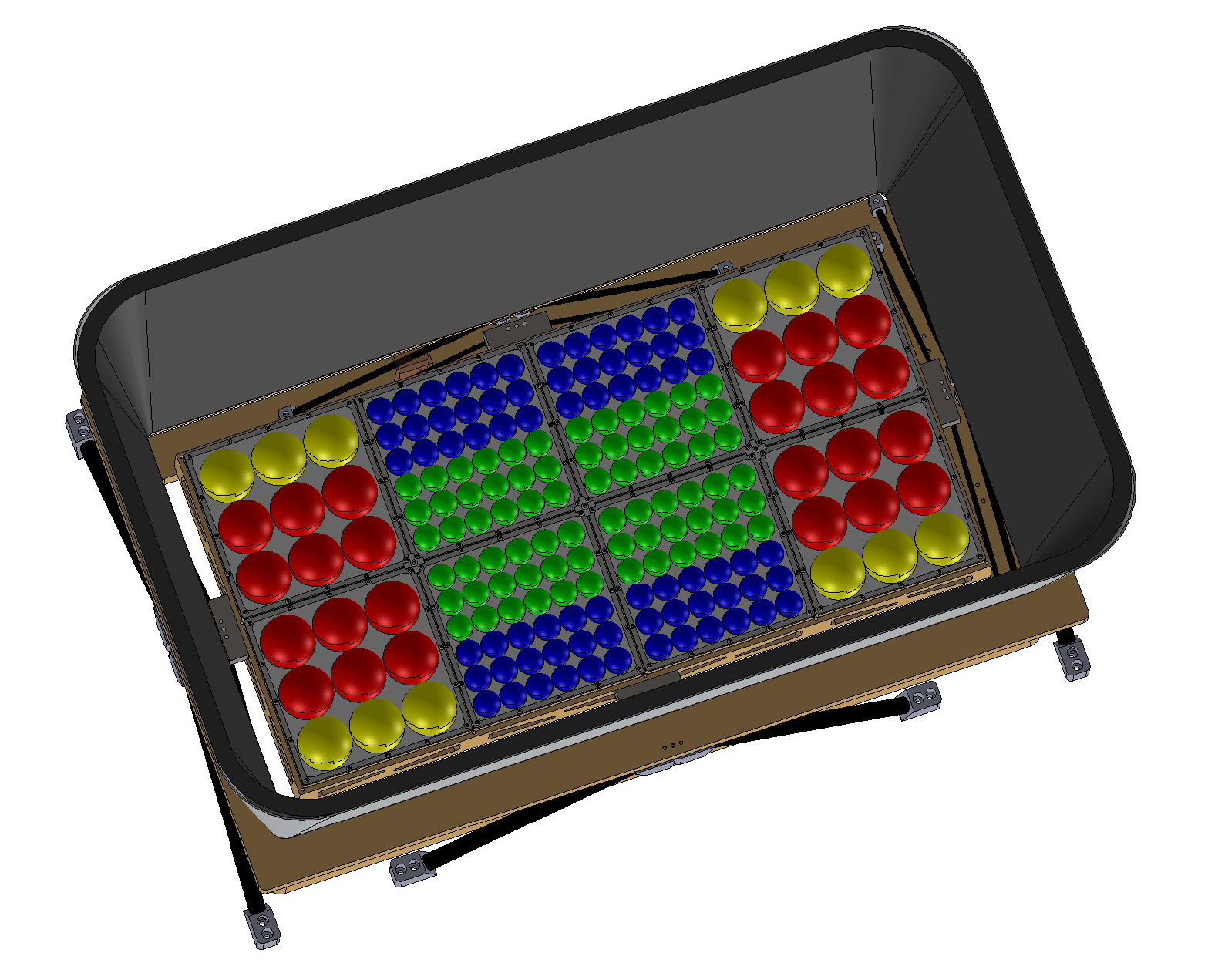}}
\subfigure[]
{\label{subfig:lf4Fpm}
\includegraphics[width = 0.52\textwidth]{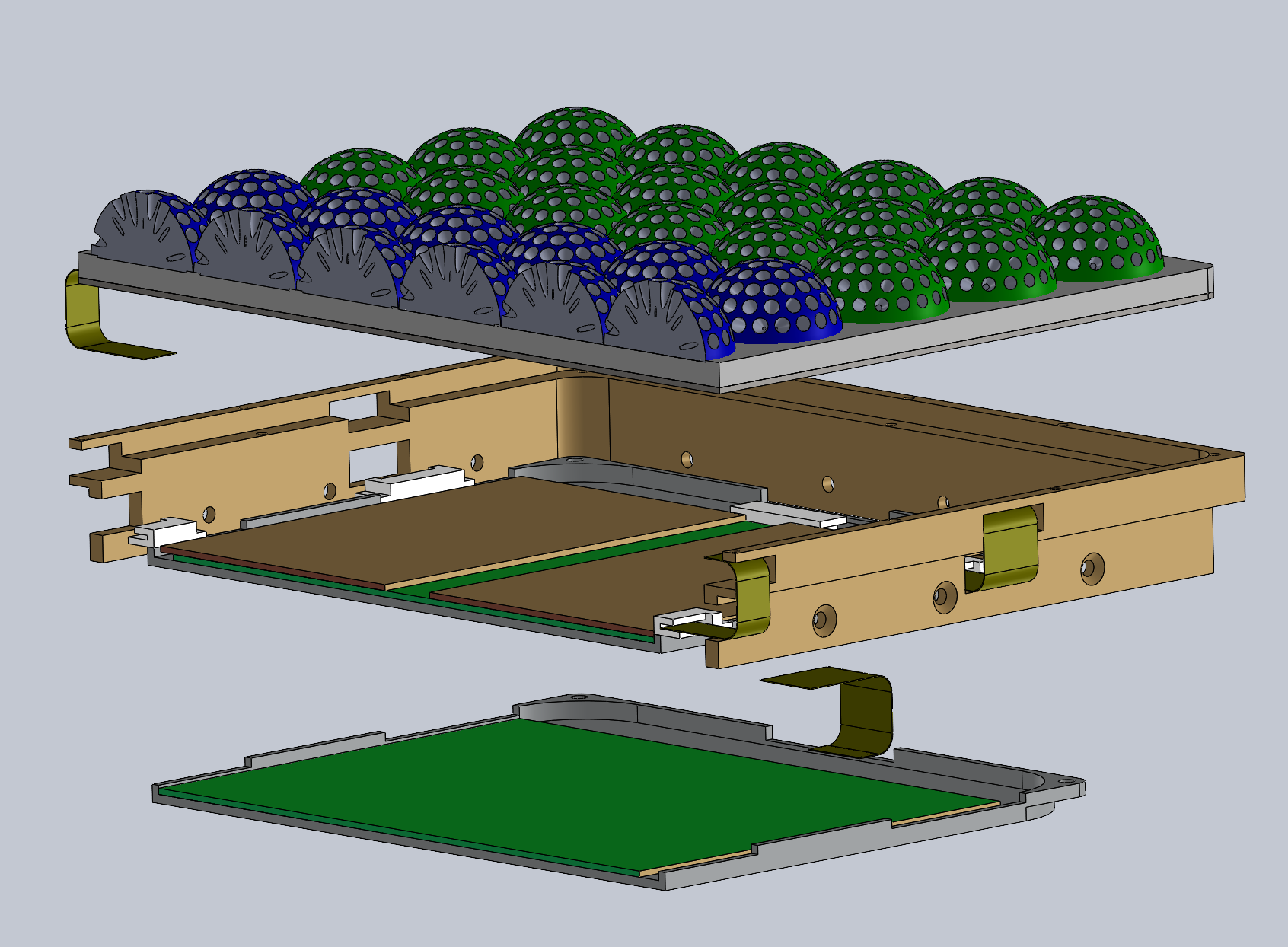}}
\caption{(a) A CAD rendering of the \gls{lffpu} with all 8 \glspl{fpm}.  The pixels observing shorter wavelengths (blue and green) are place at the center of the focal plane to minimize stray light effects at the edge of the focal plane. (b) A cross-sectional diagram of a LF-34 \gls{fpm}. The lenslets will have laser-drilled holes that form a broadband \gls{ar} surface described in Section \ref{ssec:marsFabrication}.  Niobium bias lines carry \gls{tes} signals to the edge of the wafer where they are bonded to flexible circuitry and connected to the \gls{cr}. \label{fig:lfFpu}}
\end{figure}

These proceedings will focus on the elements of the detection chain once the light from the telescope is focused the detectors in the focal plane.  The \gls{lffpu}, will tile \glspl{fpm} into a four by two array as shown in the left hand side of Figure \ref{fig:lfFpu}.  A full description of detection chain can be found in \gls{ptep} \cite{LiteBIRD:2022_PTEP}. There are three major components in the \glspl{fpm}; the lenslets which couple the antenna to the telescope, the detectors which filter and sense the incoming photons, and the \gls{dfmux} readout electronics housed on the back (non-sky) side of the focal plane. The major elements are shown in the right hand side of Figure \ref{fig:lfFpu}. This technology is able to reach the instantaneous sensitivity required by the \lb\ mission and is a mature technology with well-established use in ground-based and sub-orbital \gls{cmb} experiments~\cite{Arnold:2010_POLARBEAR, Westbrook:2018_SimonsArray, Carter:2018_SPT3G2year}.

\subsection{Lenslets}
\label{ssec:lenslets}

A `lenslet' is a contact lens (usually silicon or alumina) concentrically aligned to a planar antenna to increase the forward gain of said antenna with minimal absorption loss \cite{Suzuki:2013_Thesis, Edwards:2012_LensletSinuous}. In the \lb\ design, the lenslets are individually fabricated silicon hemispheres which are bonded to a silicon extension wafer.  The hemisphere combined with this extension wafer approximates an ellipse to optimize the forward gain.  This technology requires \gls{ar} coating or surface to minimize the loss at the vacuum-dieletric interface \cite{Farias:2022_MARS}.  \lb\ will implement a \gls{mars} on all of the lenslets in the \gls{lft} and \gls{mft}.  The beam properties of the detectors are primarily determined by the diameter and symmetry of the lenslet and its \gls{ar} surface rather than the antenna itself.  A summary of pixel properties with their beam sizes is given in Table \ref{tbl:lowFrequencyBands}.

The LHS of Figure \ref{fig:singlePixelCoupling} shows a \gls{hfss} simulation setup of a lenslet placed over a LF-4 pixel. In this simulation the \gls{ar} coating is simulated as a traditional three layer quarter-lambda \gls{ar} coating.  In the case of \lb\, the lenslet will appear as they do in the RHS of \ref{fig:singlePixelCoupling}.

\subsubsection{Metamaterial Anti-Reflection Surfaces (MARS)}
\label{sssec:marsLenslet}

\begin{figure}[ht!]
\centering
\subfigure[Rendering of an HFSS LF-4 pixel]
{\label{subfig:lf4Simulation}\includegraphics[width=0.44\textwidth]{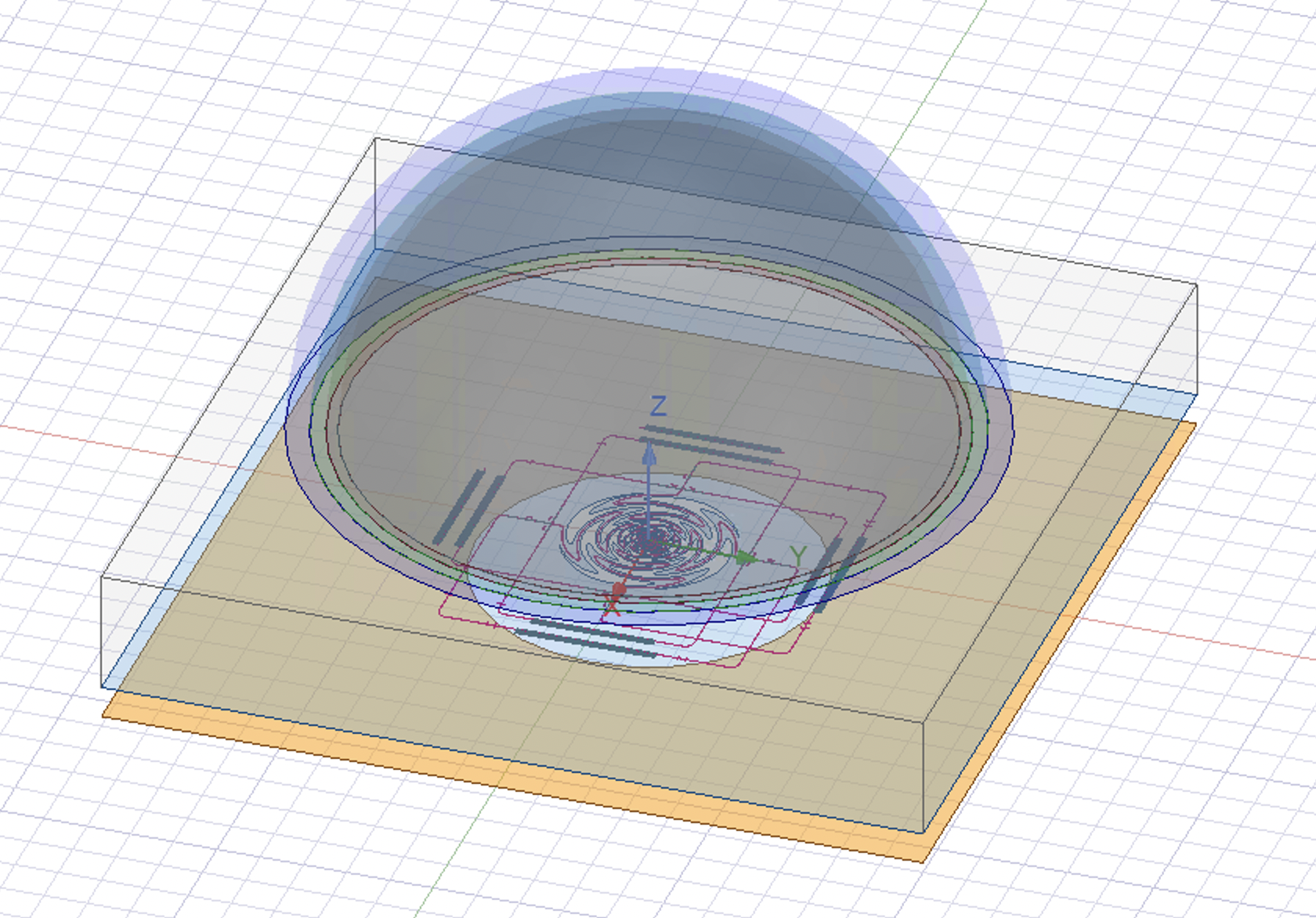}}
\subfigure[CAD of \gls{mars} on tip of lenslet]
{\label{subfig:singleLensletMars}\includegraphics[width=0.46\textwidth]{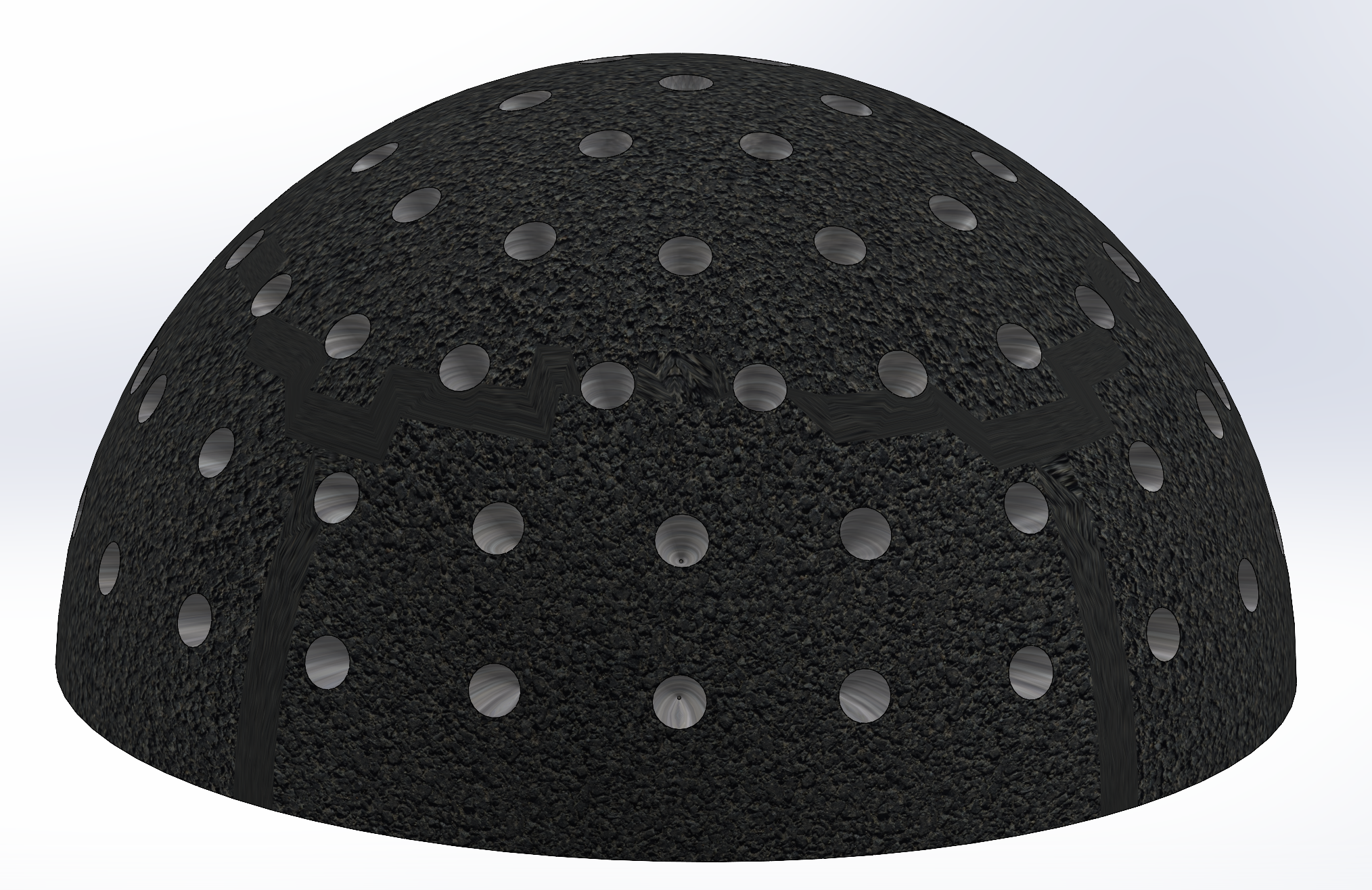}}
\subfigure[Cross section of a lenslet with a \gls{mars}]
{\label{subfig:singleLensletMarsCrossSection}\includegraphics[width=0.75\textwidth]{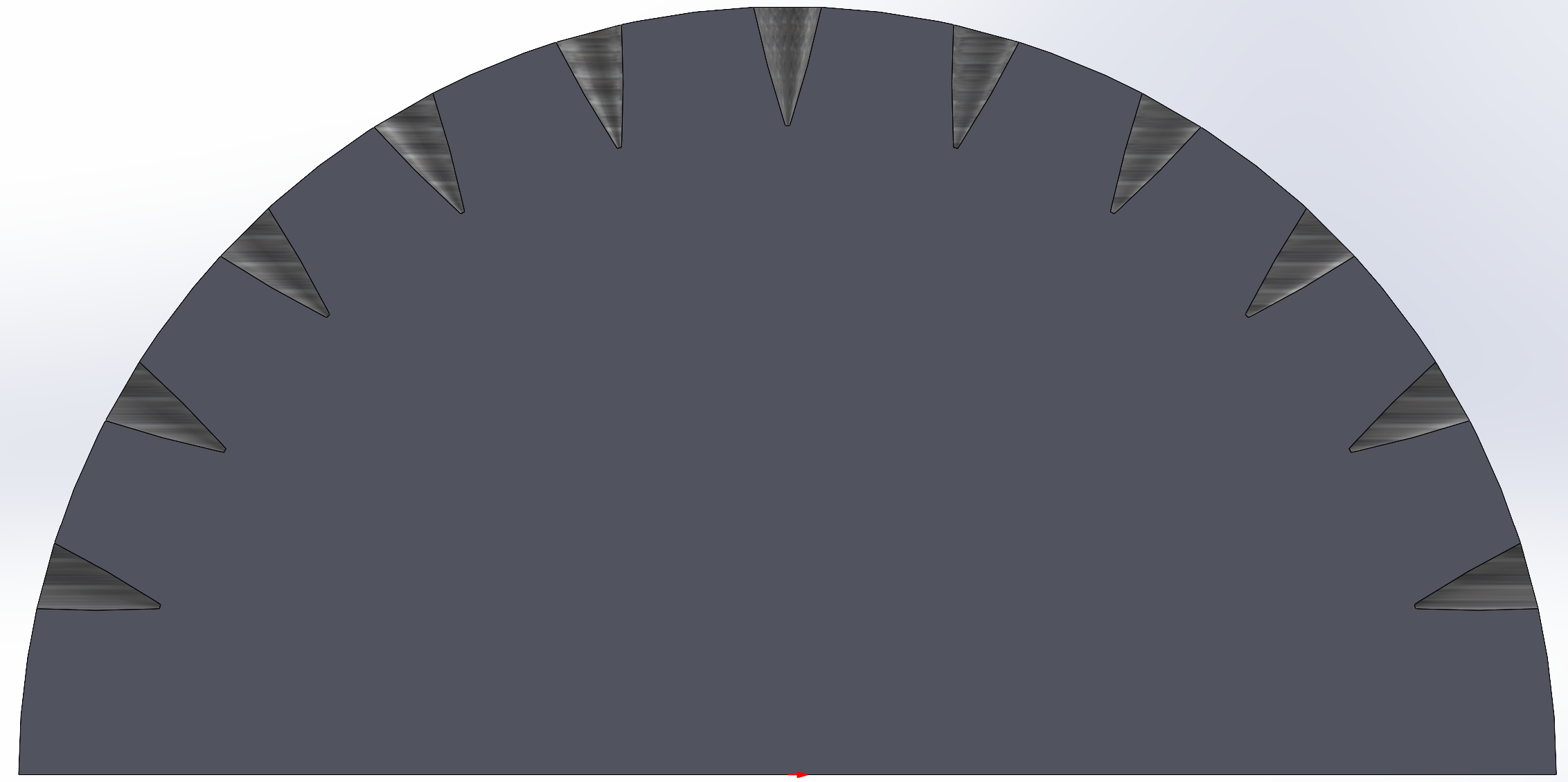}}
\caption{The coupling architecture for a single LF-4 pixel.  (a) shows a sinuous antenna behind a 16~mm diameter lenslet in \gls{hfss}.  The \gls{mars} is approximated with using a three layer \gls{ar} coating.  The yellow Pd backshort has a 7.5~mm diameter hole to let radiation through to excite the antenna. (b) is a \gls{cad} drawing of a lenslet with a hypothetical \gls{mars} laser-diced into its surface for demonstration purposes.  The deployment lenslets will have holes much much smaller than the diameter of the lenslets. (c) is rendering of a cross section of a portion of a lenslet with a laser-diced \gls{mars}. \label{fig:singlePixelCoupling}}
\end{figure}

Metamaterials are an attractive option for \lb\ as they can be tuned to cover the entire frequency range of the \gls{lft} and \gls{mft} and do not rely on epoxies or other glues to adhere to the silicon lenslets, which can be prone to differential thermal contractions.  The metamaterial creates an effective index of refraction by strategically removing silicon from the substrate to create a sub-wavelength structure in the silicon. The shape and size of the structure determines how the index of refraction will change as function of depth in the hole itself as described in Farias 2022 \cite{Farias:2022_MARS}.  An example \gls{cad} cross section of a silicon lenslet with a \gls{mars} is shown in Figure \ref{fig:singlePixelCoupling}.

\subsection{Planar Sinuous Antenna TES Bolometer Arrays}
\label{ssec:planerSinuousAntennaTesBolometerArras}

In a single pixel, once the radiation from the telescope excites the planar sinuous antenna, it couples to low-loss superconducting micro-strip transmission lines, which carry the photon through multiple lumped element pass-band and low-pass filters, multiple cross-unders for orthogonal polarizations and bands, and \gls{rf} termination at the bolometer islands.  These elements are shown in Figure \ref{subfig:singleLf4Pixel}.

\subsubsection{Sinuous Antenna}
\label{sssec:sinuousAntenna}

Detailed descriptions, including important systematics of sinuous antennas, can be found in previous reports \cite{LiteBIRD:2022_PTEP, Westbrook:2020_SPIE}.  We chose this technology for the \gls{lft} and \gls{mft} as its broadband nature is well suited to cover the frequency ranges required by \lb.  For this application we describe the specific antenna used to fabricate a LF-4 prototype as shown in Figure~\ref{fig:singlePixelCoupling}.

\subsubsection{TES Bolometers}
\label{sssec:tesbolos}

The designs for the \gls{tes} bolometers are quite similar for each band due the fact that they share a common operating temperature and \gls{cr} system.  Small variations in expected optical loading will lead to different thermal conductance in each of the bolometers as the \glspl{fpu} all operate at the same \gls{tb}. The \gls{tc} the bolometers will be $\sim$~170-200~mK the \gls{rn} will be $\sim$1~$\Omega$.  Once the conductance is tuned, the heat capacity of each bolometer is adjusted so that it has the same \gls{t0} of 33~ms.  As shown in Table \ref{tbl:commonSpecs}, the bolometers are intended to be operated at a loopgains $\geq$10 for an operation time constant of 1.5 to 3~ms.  More common bolometric parameters are shown in Table \ref{tbl:commonSpecs} and details of each band and expected optical power are shown in Table \ref{tbl:lowFrequencyBands}.

\subsubsection{RF Filters}
\label{sssec:rffilters}

The details of the \gls{rf} components vary from pixel to pixel and band to band, but the paradigm is the same for all of the pixels as all of the arrays define the pass-bands directly on the wafer using a combination of low-loss superconducting micro-strip transmission lines, on wafer pass-band definition, cross-unders for orthogonal polarizations, and \gls{rf} termination at the bolometer island.    For \lb\ the micro-strip lines are constructed with Niobium and a low-loss SiNx layer.   Traditional \gls{rf} circuit elements are patterned to the transmission lines to create the designed passbands.   Table  \ref{tbl:lowFrequencyBands} summarizes the band centers and fractional bandwidths ($\Delta \nu / \nu$) for all of the \lb\ bands, which is typically $\sim$23\% to $\sim$30\%. 

\begin{table}[ht]
\centering
\begin{tabular}{ccccccccc}
\small

\begin{tabular}[c]{@{}c@{}}Module \\ Name\end{tabular}     & \begin{tabular}[c]{@{}c@{}}Pixel \\ Name  \end{tabular} & \begin{tabular}[c]{@{}c@{}}Pixel Size \\ ($mm$) \end{tabular} & \begin{tabular}[c]{@{}c@{}}Pixel \\ Count \end{tabular} & \begin{tabular}[c]{@{}c@{}}Frequency\\ ($GHz$) \end{tabular}& \begin{tabular}[c]{@{}c@{}}Fractional\\ Bandwidth \end{tabular}  & \begin{tabular}[c]{@{}c@{}}Beam Size \\ (arcmin) \end{tabular} & \begin{tabular}[c]{@{}c@{}}Optical Power \\ ($pW$) \end{tabular} \\
\hline
                          &                                        &&& \pixelred{40} &  \pixelred{0.3}  & \pixelred{70.5} & \pixelred{0.2918}  \\
                          &                                        &&& \pixelred{60} &  \pixelred{0.23} & \pixelred{51.1} & \pixelred{0.2419}  \\ 
                          & {\multirow{-3}{*}{\pixelred{LF-1}}}    & {\multirow{-3}{*}{\pixelred{32}}}  & {\multirow{-3}{*}{\pixelred{24}}}& \pixelred{78} &  \pixelred{0.23} & \pixelred{43.8} &  \pixelred{0.2686}  \\  
                          &                                        &&& \pixelyellow{50} &  \pixelyellow{0.3} & \pixelyellow{58.5} & \pixelyellow{0.306}   \\
                          &                                        &&& \pixelyellow{68} &  \pixelyellow{0.23} & \pixelyellow{47.1} & \pixelyellow{0.2709}  \\
{\multirow{-6}{*}{LF12}}  & {\multirow{-3}{*}{\pixelyellow{LF-2}}}  & {\multirow{-3}{*}{\pixelred{32}}} & {\multirow{-3}{*}{\pixelyellow{12}}} & \pixelyellow{89} &  \pixelyellow{0.23} & \pixelyellow{41.5} & \pixelyellow{0.2419} \\                         
                          &                                        &&&\pixelblue{68} &  \pixelblue{0.23} & \pixelblue{41.6} & \pixelblue{0.3279}\\
                          &                                        &&&\pixelblue{89} &  \pixelblue{0.23} & \pixelblue{33.0} & \pixelblue{0.3163} \\
                          & {\multirow{-3}{*}{\pixelblue{LF-3}}}    & {\multirow{-3}{*}{\pixelblue{16}}} & {\multirow{-3}{*}{\pixelblue{72}}} & \pixelblue{119}        &  \pixelblue{0.3} & \pixelblue{26.3} & \pixelblue{0.3765}   \\                              
                          &                                        &&& \pixelgreen{78} &  \pixelgreen{0.23} & \pixelgreen{36.9} & \pixelgreen{0.3272} \\
                          &                                        &&& \pixelgreen{100} &  \pixelgreen{0.23} & \pixelgreen{30.2} & \pixelgreen{0.3071}\\
 {\multirow{-6}{*}{LF34}} & {\multirow{-3}{*}{\pixelgreen{LF-4}}}   & {\multirow{-3}{*}{\pixelgreen{16}}} & {\multirow{-3}{*}{\pixelgreen{72}}} & \pixelgreen{140} &  \pixelgreen{0.23} & \pixelgreen{23.7} & \pixelgreen{0.3557}  \\                            
\end{tabular}
\caption{All of the bands in the \gls{lffpu} listed by pixel name, size, and count including each bands beam size and expected optical load.  Measurements of prototype bolometers are shown in \ref{sec:testing}.}
\label{tbl:lowFrequencyBands}
\end{table}

\begin{table}
    \centering
    \begin{tabular}{l|l}
         Design & Goal \\
         \hline 
         \hline
         Pixel in-band optical efficiency & $\geq$ 80\% \\
         \gls{psat} & 2-3X optical power \\ 
         On-sky end-to-end yield & $\geq$ 80\%\\
         \gls{fpu} \gls{tb} & 100-120~mK \\
         \gls{tc} variation & $\leq$ 7\% \\
         \gls{tes} \gls{rn} & 1.0~$\Omega$ \\
         Operating \gls{tes} resistance & 0.6 to 0.8 $\Omega$ \\
         Parasitic series resistance & 0.05 to 0.2 $\Omega$ \\
         \gls{t0} & 33~ms \\
         Loopgain during operation & $\geq$~10 \\
         Common 1/f-knee & $\leq$ 20~mHz\\
         \gls{fpu} lifetime & $\geq$ 3 years \\
         \hline
    \end{tabular}
    \caption{A summary of the common  optical and bolometric design goals of the \lb\ detectors.  We expect that there will be little deviation from these goals during the development of the \glspl{fpu} for \lb. \label{tbl:commonSpecs}}
\end{table}

\subsubsection{Cosmic Ray Mitigation}
\label{sec:cosmicrays}

Following the lessons learned from the Planck satellite mission regarding the importance of mitigating cosmic ray events, our group has spent a considerable effort to building in mitigation techniques directly into both our detector and module designs \cite{Miniussi:2014_Planck_CosmicRays}.   At the detector level we have built prototypes with mitigation directly at the thermal isolation site of the bolometer as shown in Section ~\ref{sec:testing}.

\begin{figure}[ht!]
\centering
\subfigure[Photograph of a bolometer with frontside cosmic ray mitigation features]
{\label{subfig:lf4Simulation} \includegraphics[width=0.43\textwidth]{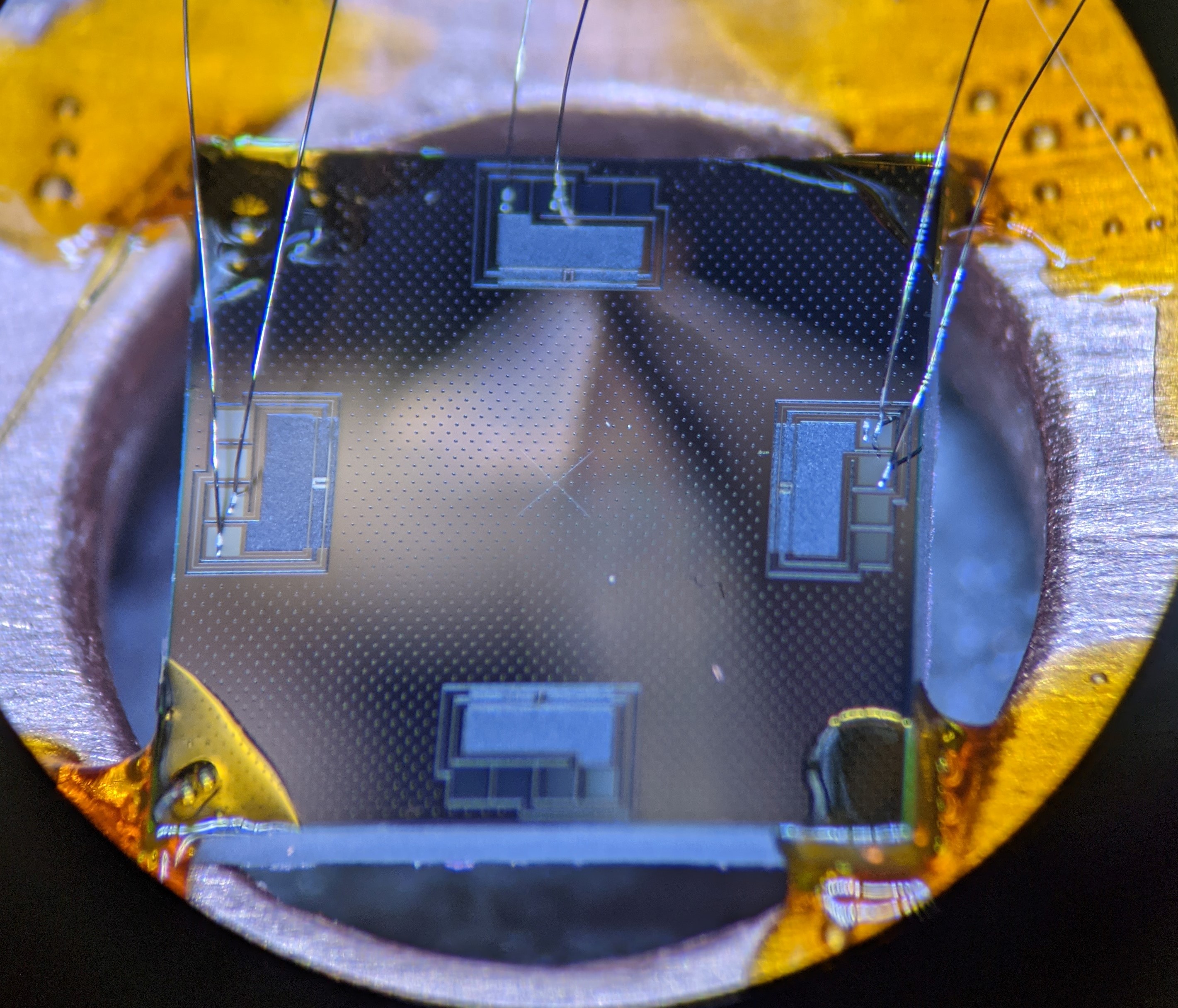}}
\subfigure[Photograph of a test chip with Au bonds made to the backside coated with Pd]
{\label{subfig:lf4Simulation} \includegraphics[angle=90, width=0.5\textwidth]{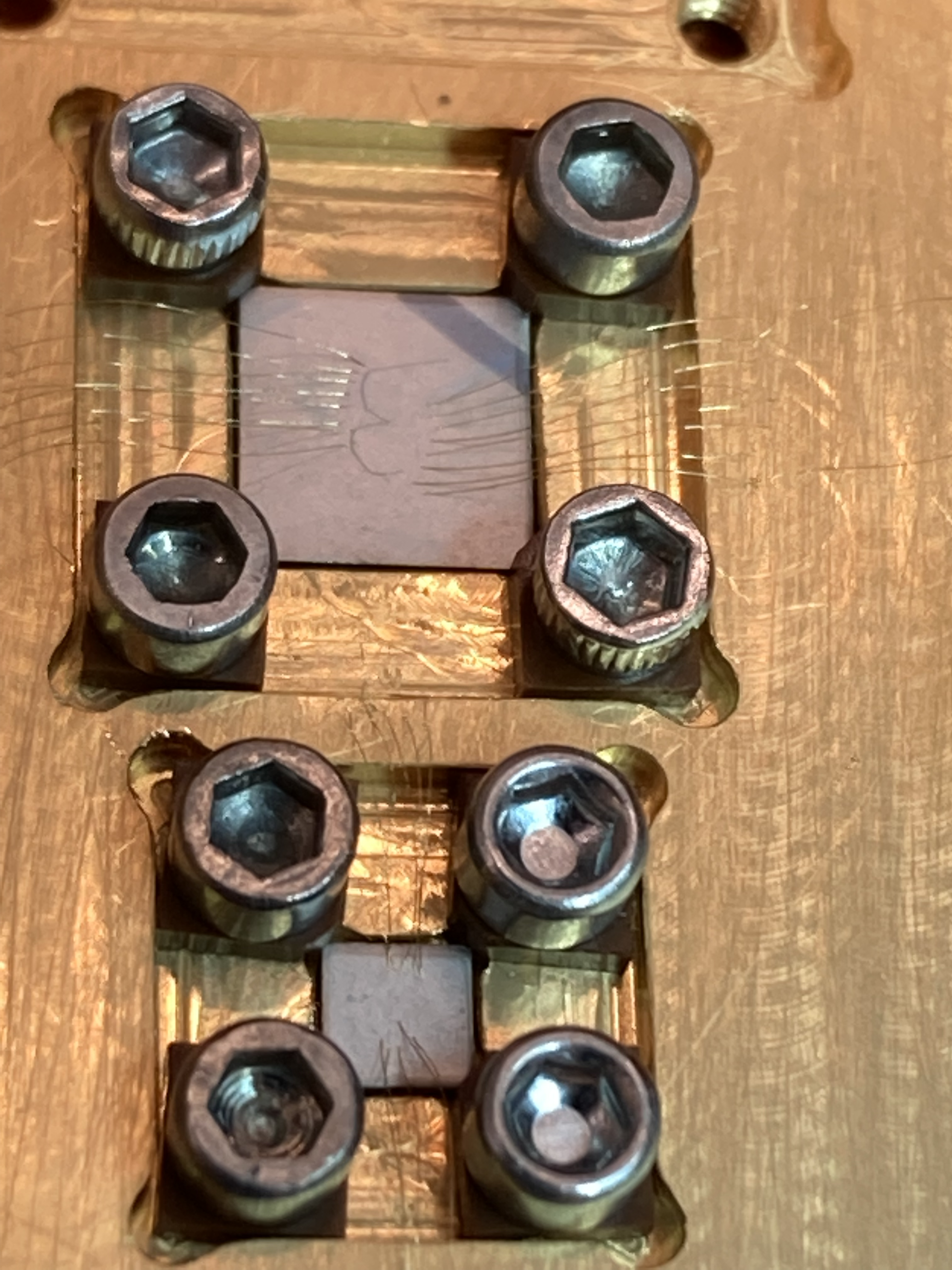}}
\caption{We are optimizing the overall coupling design to \textit{not} couple well to cosmic ray events that will strike the \glspl{fpm} during flight.   The LHS features a photograph of a bolometer designed to disrupt the flow of phonons generated by \gls{cr} events on the surface of the wafer.  The RHS is a photograph of the backside of a test chip coated with Pd with added Au wirebonds to improve thermal conduction between the device wafer to the invar frame. \label{fig:cosmicRays}}
\end{figure}

For \lb\ bolometers will be thermally isolated and the wafers will be thermally sunk with a normal metal, such as Pd. Initial tests at \gls{ucb} have shown these hardware implementations to work well \cite{Westbrook:2020_SPIE}.

\subsection{Interface to Cryogenic Readout}
\label{ssec:cyrogenicreadout}

The same niobium layer that forms the micro-strip layer is also used to route the \gls{tes} bias/readout leads to the edges of the wafer to interface with the readout. For the \gls{lf} \glspl{fpm} wiring density is low compared to ground-based arrays allowing for all of the \glspl{tes} to be readout on a just two sides allowing for more flexibility with the design of the \gls{cr} interface.  For the \gls{mf} and \gls{hf} \glspl{fpm}, the wires are routed to all of the edges of the wafer for bonding. 

We are minimizing the contribution of the \gls{cr} to the noise of the detection chain.   The design principles are similar to that of \gls{sa} and \gls{spt3g} where flexible circuitry makes electrical connections from the device wafer to lithographed inductor-capacitor (``LC'') chips.  In previous experiments, a stripline which is up to a full meter in length sends the signals to \gls{sq} arrays at 4~K.  \lb\ has explored many options for how to use \gls{sq} amplifiers at various cold stages to meet the noise requirements of \lb.  To reduce the overall readout noise, the \glspl{sq} is integrated directly into the \glspl{fpm}.  This allows operation of the \glspl{sq} at colder temperatures and the ability to use shorter flexible circuitry for the initial step in the detection chain.  A full detailed discussion of the development of the \gls{cr} for \lb\ can be found in Russell 2022\cite{Russell:2022_SPIE}.

\section{Fabrication of LF-4 Prototypes}
\label{sec:fabrication}

It will be the responsibility of \gls{ucb} to fabricate the eight \gls{lf} wafers as well as all of lenslet arrays for the \gls{lffpu} and the \gls{mffpu}. In this section, we report on the fabrication status for the \gls{mars} for the lenslets and the detector arrays for the 70/100/140 (LF-4) band.

\subsection{MARS Fabrication}
\label{ssec:marsFabrication}

\begin{figure}[ht!]
\centering

 \includegraphics[width=0.95\textwidth]{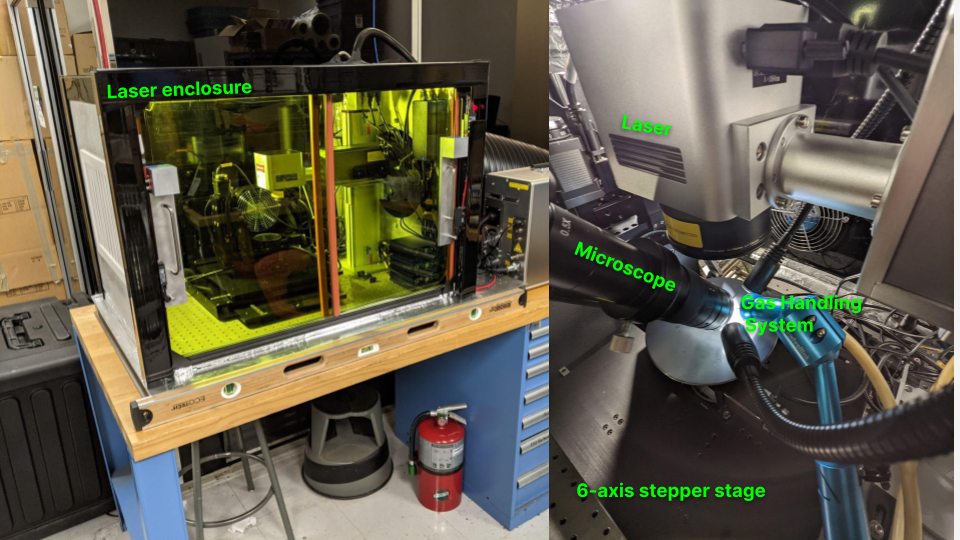}
\caption{Photographs of the laser system built at \gls{ucb} to manufacture a laser etched \gls{mars} for lenslets for \lb.  The laser enclosure houses the laser and 6 axis system to create uniform \gls{mars} on the surface of \lb\ lenslets \label{fig:laserSetUp}}
\end{figure}

The \gls{mars} for the lenslets will be fabricated at \gls{ucb} using a custom lasing-system.  This system is comprised of a laser enclosure, the laser itself, which includes a galvanometer for precise laser scanning, a gas-handling system for process control, in-situ real time video microscopy, and a computer controlled 6-axis stepper motor system for precision placement of the lased holes as shown in Figure~\ref{fig:laserSetUp}.   This system is capable of producing sub-wavelength structures relevant to all of the lenslet-coupled bands in \lb \cite{Farias:2022_MARS}.  

\subsubsection{Laser Enclosure}
\label{sssec:laserEnclosure}

\begin{figure}[ht!]
\centering
\subfigure[Programmed Sites]
{\includegraphics[width=0.49\textwidth]{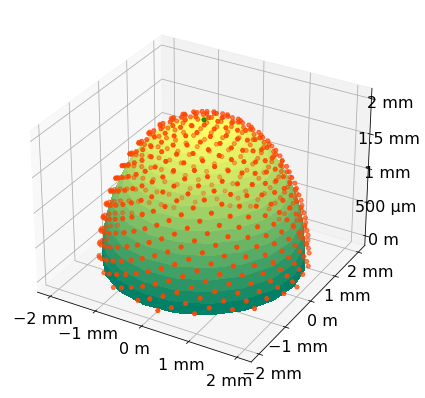}}
\subfigure[Lased Sites]
{\includegraphics[width=0.49\textwidth]{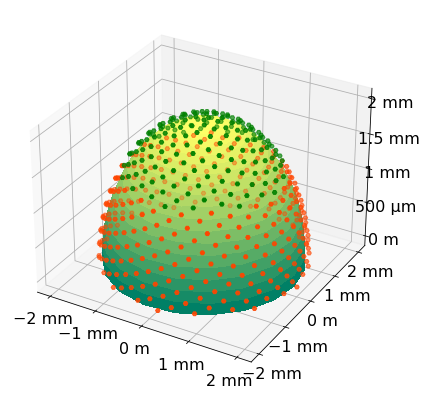}}
\caption{(a) A software rendering of hole locations to be lased by the laser system.  As described in Section \ref{sssec:laserEnclosure}, the 6-axis table is cable of precisely placing a 3D object in an arbitrary location beneath the fixed laser.  (b) As the software executes the fabrication program the computer commands 6-axis stepper system to precisely place the lenslet beneath the laser beam and marks the hole as lased with a green dot.  \label{fig:softWareControl}}
\end{figure}

The laser enclosure consists of all of the necessary Class-1 safety precautions to provide researchers with proper safety during the laser-dicing process.   The laser operates at 1064~nm and is capable of pulse-widths down to 2~ns, repetition rates up to 4~MHz, and powers up to 100~W with a minimum spot size of 34~$\mu$m.  Critically, this spot size is smaller than the largest feature size required for \lb\ \gls{mars}. Traditionally, laser machining of silicon requires femtosecond to picosecond pulse widths to avoid melted silicon and oxide production. The required structure is achieved with a relatively long, nanosecond pulse width by combining fast scanning of the laser galvanometer over the silicon surface with a gas handling system to actively clear silicon and reduce oxide production.  This greatly improved the fidelity of pattern transfer of the desired to achieved shapes of the lased holes as shown in Figure \ref{fig:lasedHolesTest}.  The gas handling system holds two high pressure nozzles at fixed locations that continuously flow high pressure nitrogen at the lasing site, as well as a third nozzle placed to remove residue across the surface of the sample.  We also installed an in-situ microscope and video camera system to have real-time monitoring of the laser ablation which has provided valuable feedback to the manufacturing process.  The final component of the laser system is the 6-axis stepper motor system with a step size of 0.625~$\mu$m.  This system will translate and rotate individual hemispheres underneath the lasing site to create the uniform arrays of holes on the surface. Software can generate arbitrary patterns on 3D surfaces to create any type of geometry. Figure \ref{fig:softWareControl} shows hypothetical locations to be lased by the laser on the surface of a hemisphere of silicon.  Each red dot specifies a location on the lenslet surface that will be lased by manipulating the lenslet in 3D space underneath the laser.

\subsubsection{Future Development of MARS}
\label{sssec:futureDevelopmentOfMars}

\begin{figure}[ht!]
\centering
\subfigure[Photograph of 150 GHZ \gls{mars}] {\label{fig:mars}\includegraphics[width=0.41\textwidth]{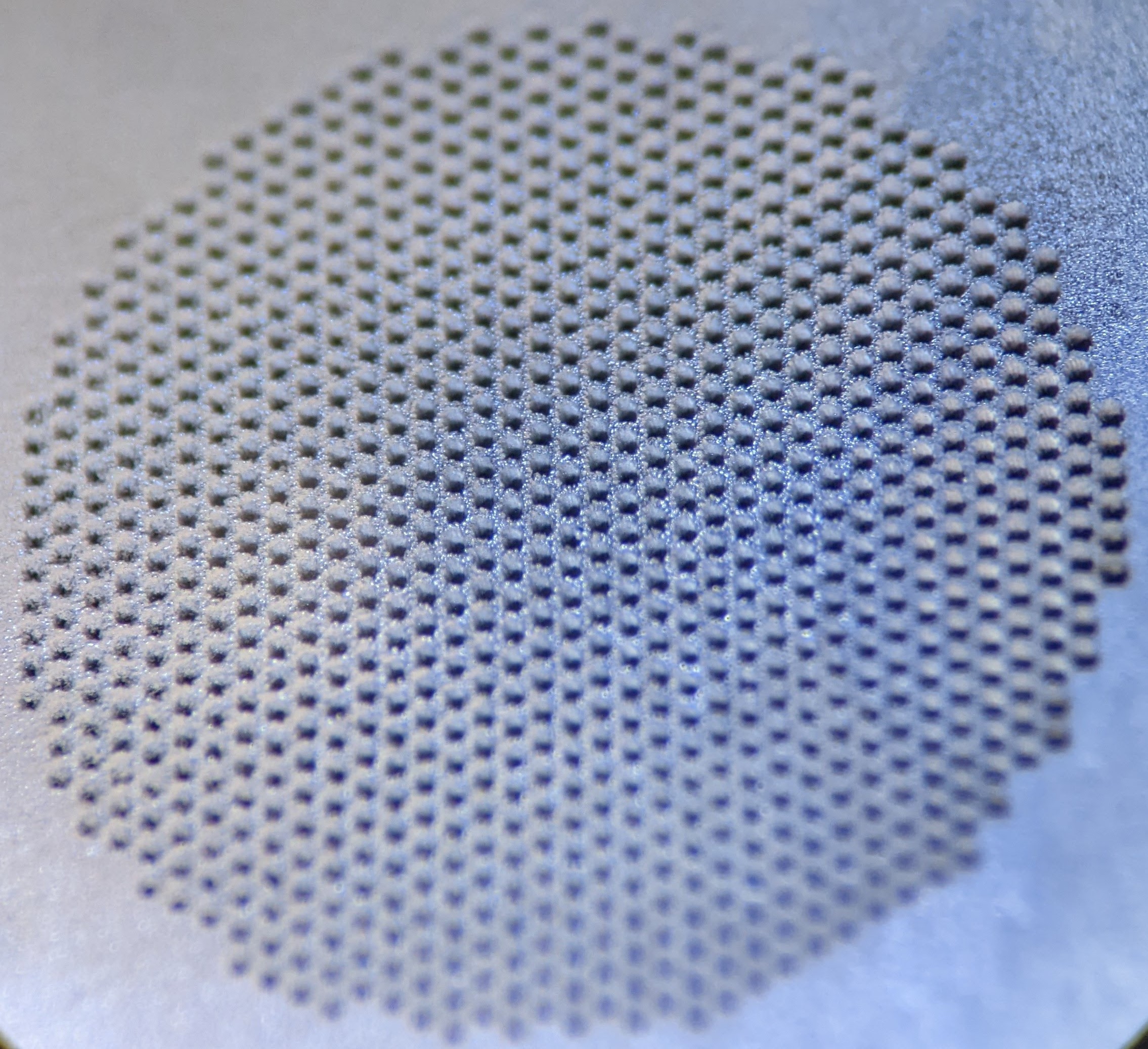}}
\subfigure[Confocal Microscopy Scan] {\label{fig:confocalMicroscopy}\includegraphics[width=0.53\textwidth]{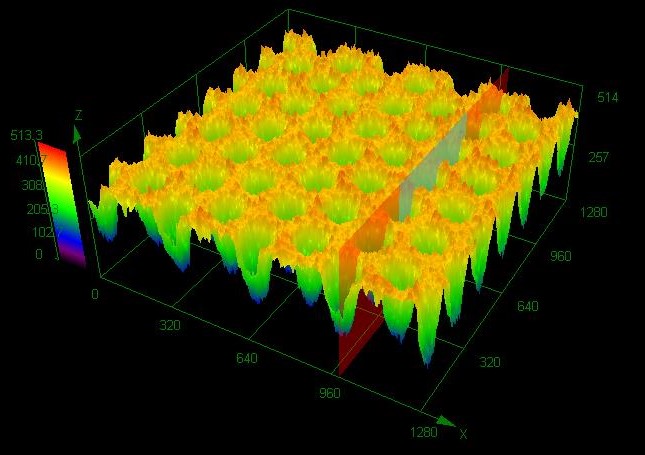}}
\caption{(a) Video 1. Photograph of a 150~GHz \gls{mars} laser-ablated into a silicon surface. (b) A confocal microscopy scan of the holes in (a).  In addition to excellent hole-to-hole uniformity, there is good fidelity between the intended and achieved geometries of the individual holes. http://dx.doi.org/10.1117/12.2630574 \label{fig:lasedHolesTest}}
\end{figure}

As a first step in the development in this technology, we have lased a single layer coating optimized for 150 GHz into a flat silicon substrate shown in Figure \ref{fig:lasedHolesTest}.  In this demonstration, we laser-diced a \gls{mars} series of hexagonal holes in a closed-hex pattern on a silicon substrate.   Each hole is a regular hexagon with a side length of 90~$\mu$m laser-drilled to a depth of 270~$\mu$m.  There is a 30~$\mu$m gap between the edges of nearest neighbors. Photographs and confocal microscopy scans of this pattern can be seen in Figure \ref{fig:lasedHolesTest}. 

Future work will include the optimization for the seven different pixel types that will deploy lenslets with laser-ablated \gls{mars}.   The 150~GHz \gls{mars} is very similar to the most challenging case of the MF-2 band, which requires good transmission up to 184~GHz.

\subsection{Device Wafer Fabrication}
\label{ssec:deviceWaferFabrication}

The standard fabrication flow of sinuous antenna bolometers at \gls{mnl} \gls{ucb} is well described in previous technology development papers for \lb \cite{Westbrook:2018_SimonsArray, Westbrook:2020_SPIE}.  These proceedings describes the initial `pre-processing' fabrication steps we took to coat the back side of the wafers with palladium for cosmic ray mitigation.
	
\subsubsection{Backside Palladium Deposition}
\label{sssec:backsidePalladiumDeposition}
As discussed in Section \ref{sec:couplingDesign}, mitigating cosmic ray events in the bolometer timestreams is critical to producing a high quality datasets for \lb.   Simulations in COMSOL suggest that the presence of a normal metal such Pd or Au on the backside of the wafer will drastically reduce any signals in the \gls{tes} as a result of a cosmic ray striking the substrate \cite{Stever:2021_JCAP}.   This presents a technical tension between coating the backside of the wafer with a normal metal while keeping the wafer optically transparent where the light from the telescope must pass through the wafer to stimulate the sinuous antenna.  As such we have developed a \gls{crmscb} to solve this issue. 

Although most of the fabrication steps that define the LB detector wafer; namely the antennae, filters, and bolometers; are processed on the frontside of the silicon wafer, there is an important and novel pre-process that occurs before the wafer is inserted into the frontside process flow. This is the process that defines the \gls{crmscb}. The purpose of this structure is to minimize the time to recover from thermal noise caused by ballistic phonons from the impact of cosmic rays into the bulk of the silicon substrate.

For the work presented in this proceedings, a 0.5~$\mu$m Pd \gls{crmscb} film was patterned via a lift-off process using e-beam evaporation. In order to prepare the wafer for this step, the \gls{lsn} on the backside is removed to allow the \gls{crmscb} to be in direct contact with the bulk silicon. Photo-resist is spun onto the frontside \gls{lsn} film to protect it during the removal of the backside \gls{lsn} and subsequent liftoff process. A simple CF4/O2 fluorine plasma etches the backside \gls{lsn} down to the silicon substrate.

Lithography is performed on the backside of the wafer to create a lift off pattern for the holes in the Pd behind each antennas.  When a 0.5~$\mu$m film of Pd evaporated for the \gls{crmscb} the resulting film has a tensile stress of 150~MPa. Experiments with a 2~$\mu$m film yielded a tensile stress of 350~MPa.  Thicker films of Pd may present a future technical challenge due to stress build up in this film. 

Once the Pd \gls{crmscb} was patterned, a passivation layer is added to the backside of the wafer both to; (1) to prevent Pd backside contamination of the wafer chucks on the tools used for five of the subsequent metallic and ceramic chlorine and fluorine plasma etches; and (2) to protect the Pd from the final corrosive XeF2 etch used to release the bolometer detectors from the bulk silicon.

The film chosen for this passivation is chemically inert and materially durable aluminum oxide. The deposition technique used is \gls{ald} to ensure conformal coating of the backside. A 10~nm coating, deposited at 300~C for 20 minutes, was used. In testing, this thickness allowed protection from the XeF2 etch while still being thin enough to wirebond through for thermal grounding – a critical packaging element to assembling the full detector array.

\begin{figure}[ht!]
\centering
\subfigure[Single LF-4 Pixel]
{\label{subfig:singleLf4Pixel}\includegraphics[width=0.45\textwidth]{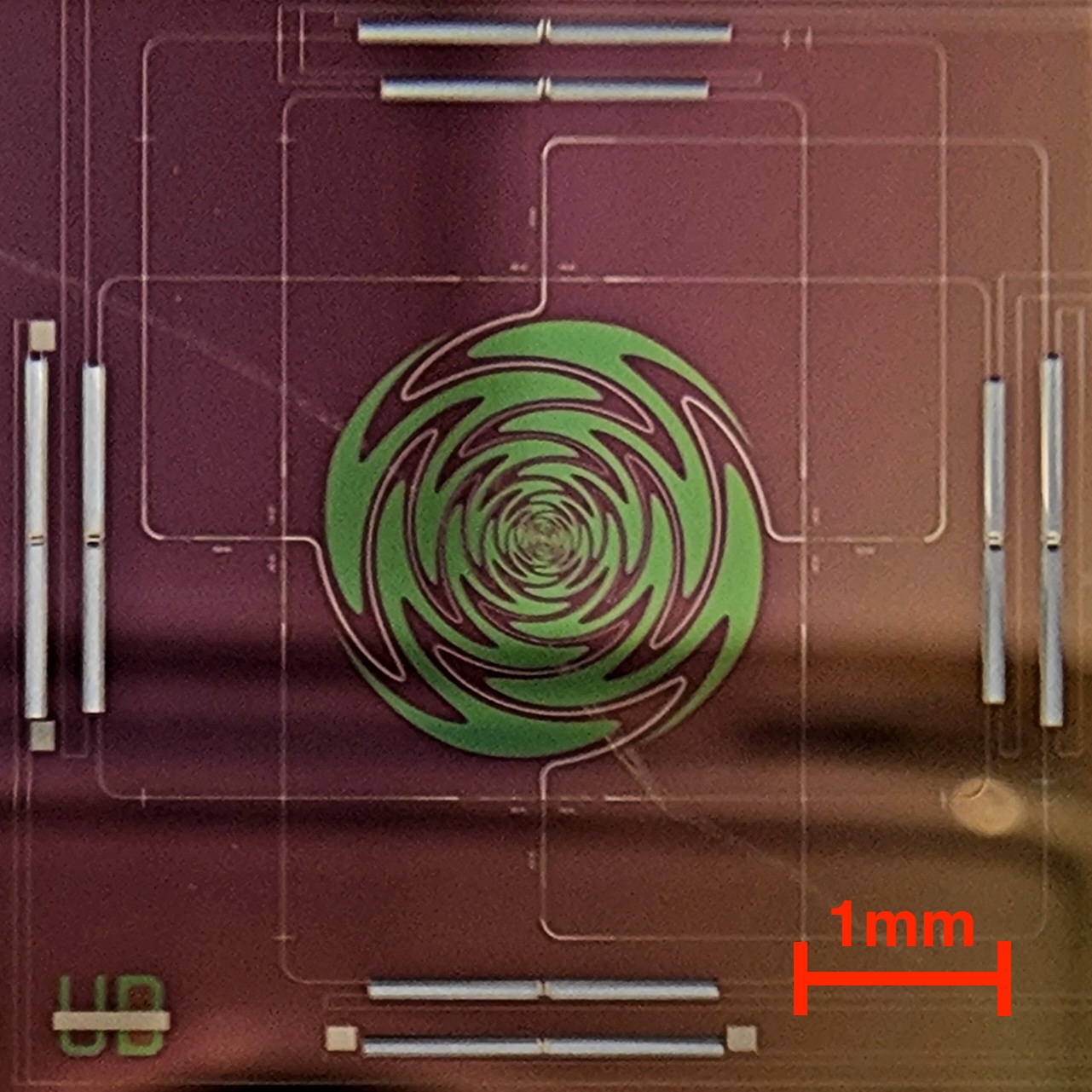}}
\subfigure[Backside coated with Pd]
{\label{subfig:singleLf4Pixel}\includegraphics[width=0.45\textwidth]{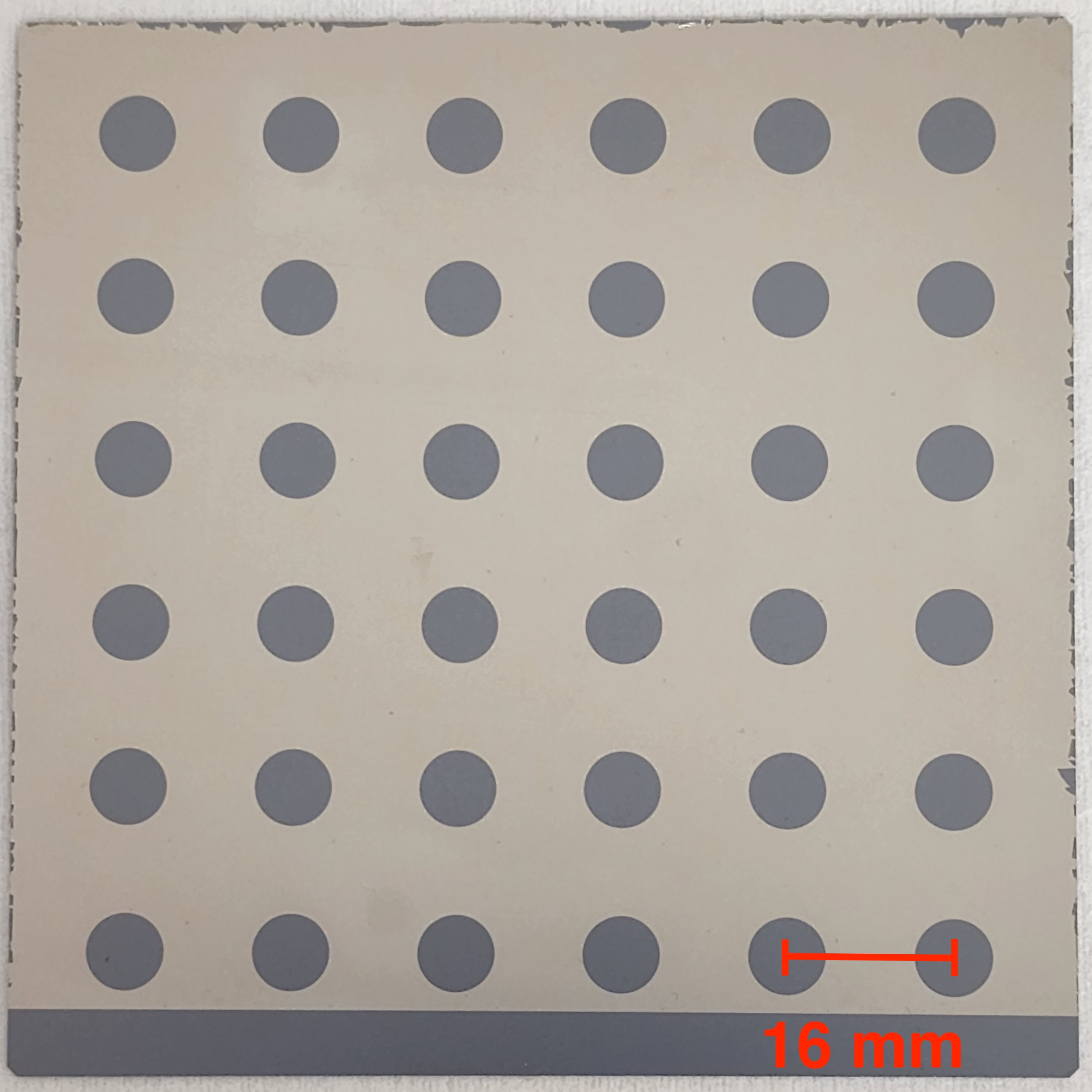}}
\subfigure[Alignment of the sinuous antenna and Pd hole] {\label{subfig:sinuousAntenna}\includegraphics[width=0.45\textwidth]{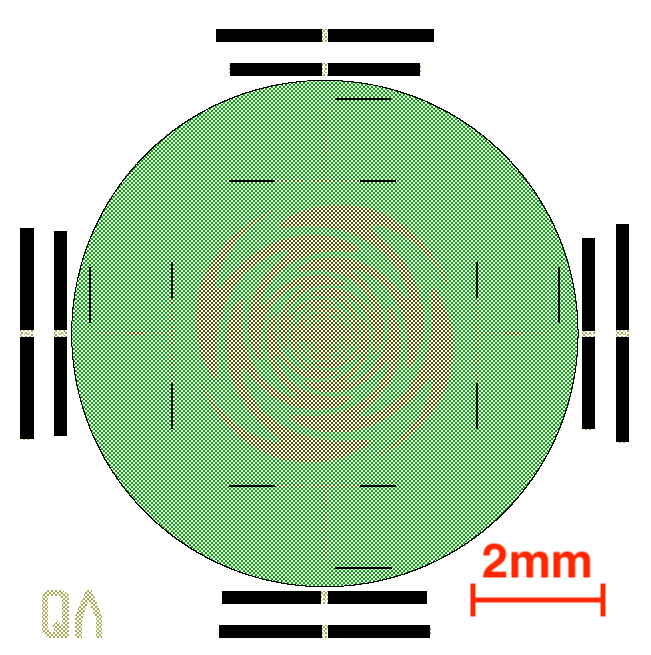}}
\subfigure[\gls{tes} Bolometer Island]
{\label{subfig:bolometerIsland}\includegraphics[width=0.49\textwidth]{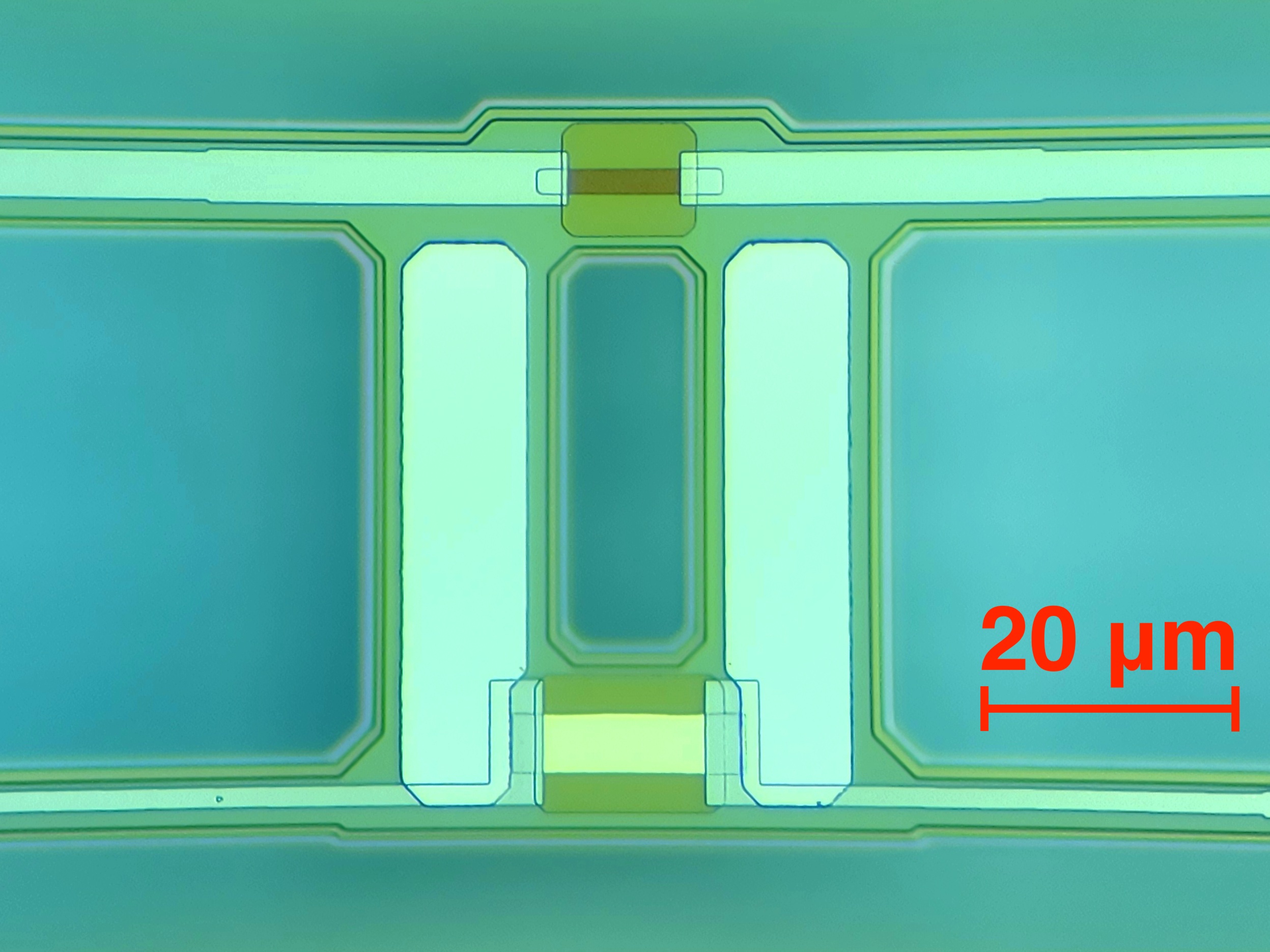}}
\caption{(a) A photograph of a full witness pixel from LF-4 prototype wafer.  Bolometric characterization data from this witness pixel can be found in Figure \ref{fig:testResults}. (b) A photographs of a LF-4 prototype wafer with backside Pd for cosmic ray mitigation.  The holes in the palladium are lithographically aligned to the antennas on the device side of the wafer to allow coupling to light from the telescope as shown in (c). (c) A CAD rendering of the sinuous antenna and bolometer placement relative to the backside Pd mitigation. The hole is place concentrically on the antenna while still directly covering the bolometers themselves. (d) A photograph of a \gls{tes} bolometer island of one of the \glspl{tes} shown in (b).  \label{fig:pixelSummary}}
\end{figure}

A prototype array with this extra Pd layer and micro-photographs of the devices on the front side is shown in Figure \ref{fig:pixelSummary}.

\subsubsection{Tc Tuning}
\label{sssec:thermalTuning}

The \lb\ mission has a baseline plan to operate the \glspl{fpu} at 100~mK with some contingency that allows for a higher base temperature.   Based on our thermal modeling of the bolometers for \lb, we find that the optimal Tc for thermal carrier noise is 1.71 times the bath temperature which minimizes the thermal carrier noise of a \gls{tes} bolometer dominated by phononic heat transport \cite{Suzuki:2013_Thesis}.  As such, a Tc range near 170~mK is targeted to meet the specifications of \lb.

To achieve this, the \glspl{tes} is thermally annealed near the end of the fabrication process before the wafer is diced.  While the exact mechanism is still to be determined, the \gls{tc} of \glspl{tes} increases monotonically when the substrate is baked at temperatures at or above $\sim$200~C.  This means that any further thermal annealing can only increase the \gls{tc}.  As such, wafers can be thermally tuned for time and temperatures that produce a \gls{tc} below the target and the thermal budget can be iteratively increased until the wafer has the desired \gls{tc}.  At this point, the wafer then is returned to the \gls{mnl} where the final fabrication steps are carried out.  This procedure ensures that we fabricate a wafer we are confident has \glspl{tes} with the correct \gls{tc}.   Figure \ref{fig:testResults} in Section \ref{sec:testing}, shows resistance versus temperature measurements from \lb\ prototype LF-4 witness pixel.

\section{Testing}
\label{sec:testing}

\gls{ucb} will characterize all of the \glspl{lffpm} in the \gls{lffpu} with support from University of Tokyo, \gls{cu}, and \gls{lbnl} \cite{Jaehnig:2022_SPIE, Russell:2022_SPIE, Ghinga:2022_SPIE}.  The primary test cryostat is a BlueFors LD400 \gls{dr} capable of cooling detector arrays to $\leq$~100~mK.  The system is currently capable of taking dark bolometric measurements to characterize \glspl{tes} from full 6 inch wafer, diced device wafers, witness pixels, and test chips.  The cyrostat houses 6 quantum design DC \glspl{sq} well suited to read out the properties of $\sim$1~$\Omega$ \lb\ \glspl{tes}.

\subsection{Bolometric Testing}
\label{ssec:bolometericTesting}

A novel feature of our fabrication and testing process is our ability to bond to an un-diced un-released device wafer.  This improves our control on tuning the \gls{tc} and \gls{rn} of the devices.  The scope of this testing is limited to measuring the stage temperature and sample resistance using four-wire resistance measurement with two LakeShore 372 AC resistance bridges.  Photographs of this testing scheme are shown in Figure \ref{fig:testBed}.

\begin{figure}[ht!]
\centering
\centering
\subfigure[Completed LF-4 Wafer]
{\label{fig:subucbdr}\includegraphics[width=0.45\textwidth]{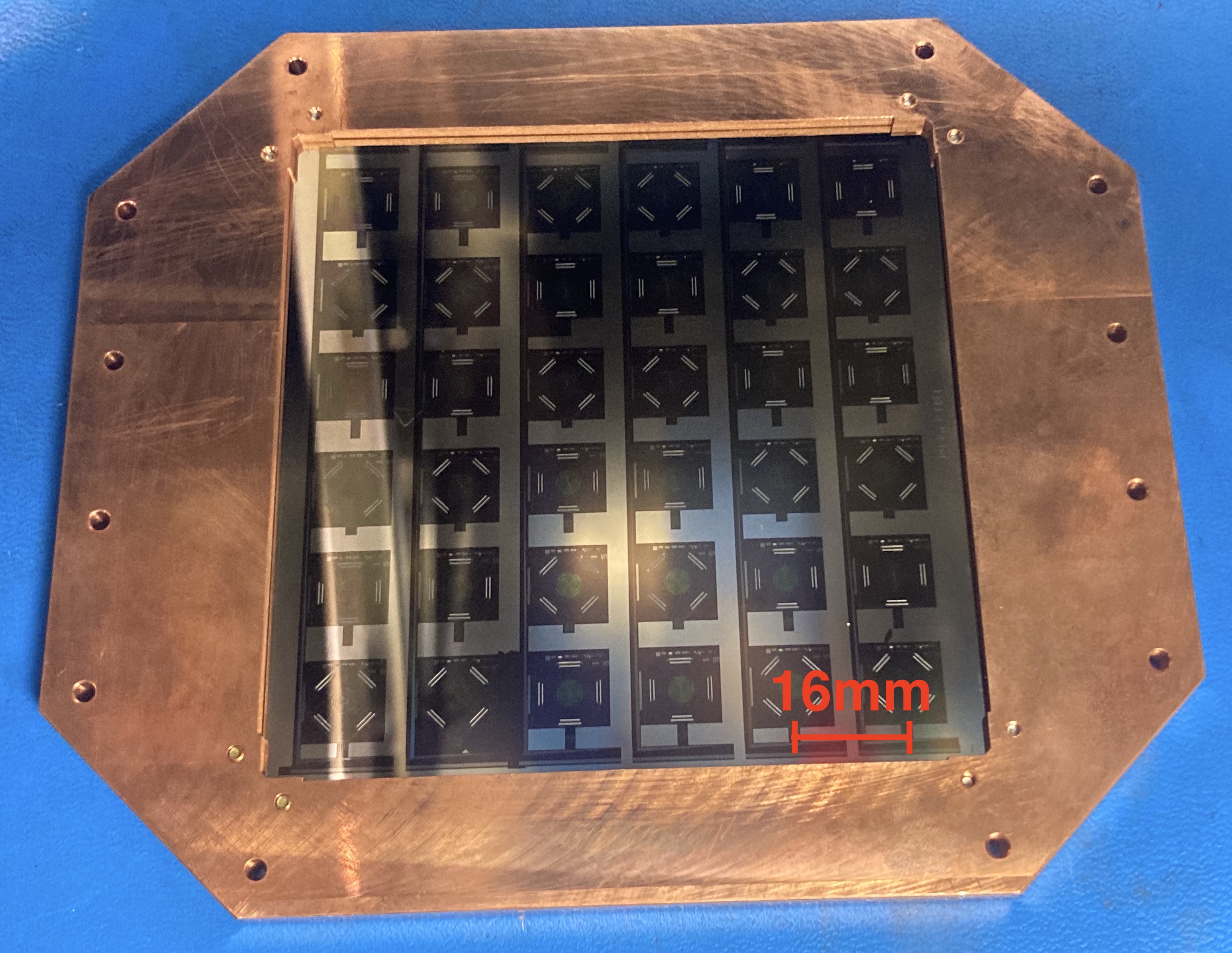}}
\centering
\subfigure[Wirebonds to the \glspl{tes} for \gls{rn} and \gls{tc} screening] {\label{fig:thermaltune}\includegraphics[width=0.46\textwidth]{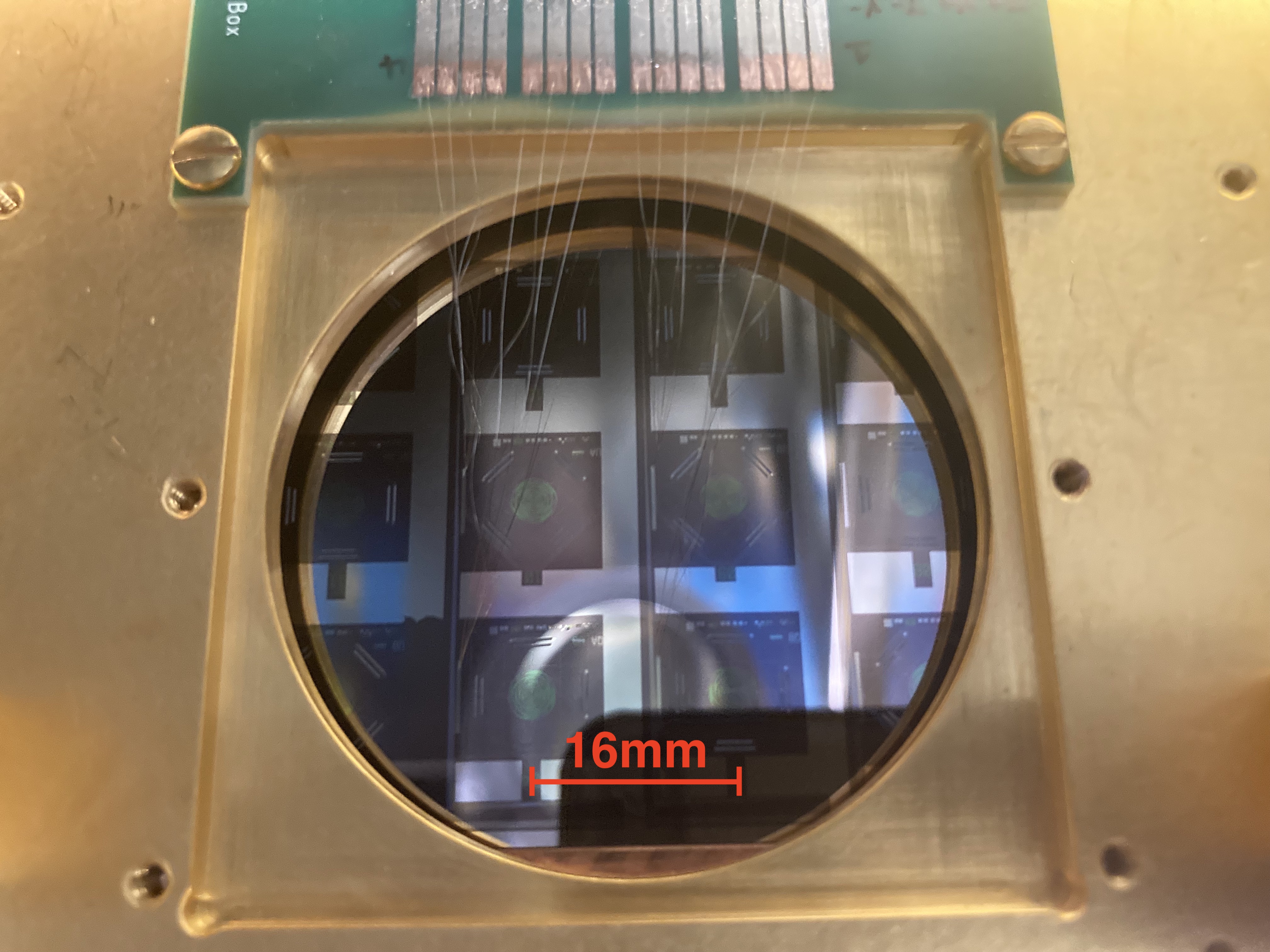}}
\caption{Example photographs of how we mount a completed wafer for testing in the \gls{dr} and \gls{ucb}.  We can measure \gls{rn} and \gls{tc} using a Lakeshore 372 AC resistance bridge and/or Quantum Design \glspl{sq}. \label{fig:testBed}}
\end{figure}

\begin{figure}[ht!]
\centering
\centering
\subfigure[RT Curve]
{\label{fig:rTCurve}\includegraphics[width=0.56\textwidth]{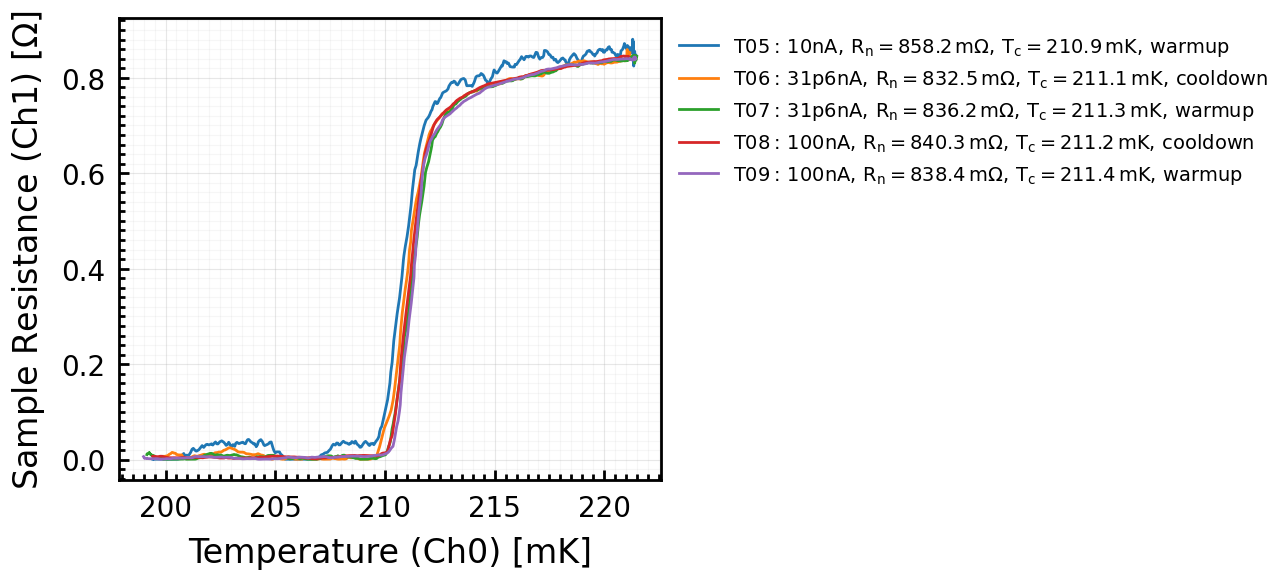}}
\subfigure[RT Curve]
{\label{fig:rTCurve}\includegraphics[width=0.4\textwidth]{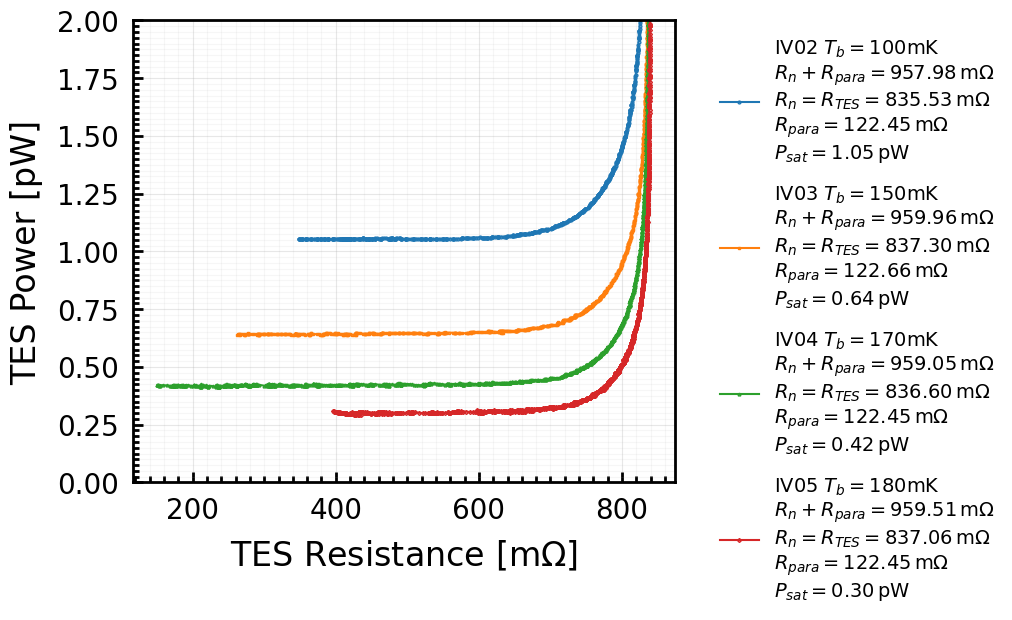}}
\centering
\caption{(a) A resistance versus temperature plot of a \lb\ witness pixel showing a \gls{tc} of 210~mK and an \gls{rn} of $\sim$~0.9~$\Omega$.  (b) I-V, R-V, and P-R curves for the same bolometer measured in (a).  The \gls{psat} of 0.9~pW which is appropriate for \lb\ as shown in Table \ref{tbl:commonSpecs}. There is good agreement between the two measurement systems.  \label{fig:testResults}}
\end{figure}

Prototype bolometers that are similar to what is expected to deploy for \lb have been fabricated and tested.  As shown in Table \ref{tbl:lowFrequencyBands} the expected optical load on all of the bolometers is 0.25 to 0.4~pW, which with a safety factor of 2.5, requires bolometers to be fabricated with \gls{psat} ranging from 0.6 to 1~pW.  Characterization of a bolometer from a prototype of a LF-4 pixel has a saturation power just above this range as shown in Figure \ref{fig:testResults}.

\subsection{Cosmic Ray Testing}
\label{ssec:cosmicRayTesting}

\gls{ucb} is testing the efficacy of the cosmic ray mitigation techniques discussed in Section~\ref{sec:couplingDesign}.  To do this, we fabricate two chips that have bolometers with the same spacing and orientation and thermal properties.  We apply mitigation to one of the chips and leave the other as a control as shown in chip Figure~\ref{fig:testBed}.   Data is logged from the chips in various radiative environments to compare the total number of events and the number of coincidence in the two chips. A detailed discussion of the experimental setup for creating a hot cosmic ray environment can be found in a previous SPIE proceedings \cite{Westbrook:2020_SPIE}.

\begin{figure}[ht!]
\centering
\centering
\subfigure[Chip Centering]
{\label{fig:subucbdr}\includegraphics[width=0.35\textwidth]{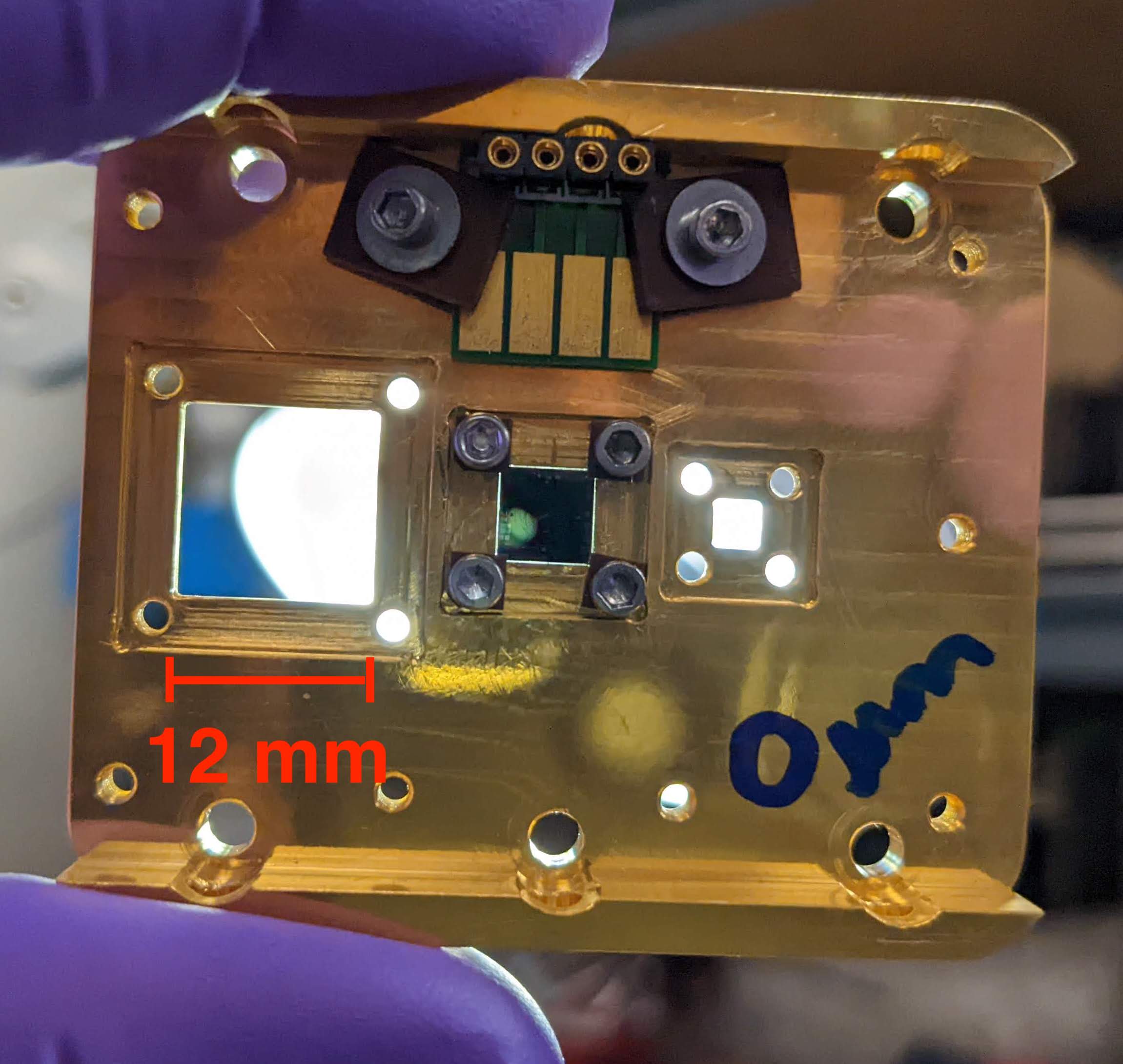}}
\centering
\subfigure[Side-by-Side AB Testing] {\label{fig:thermaltune}\includegraphics[width=0.59\textwidth]{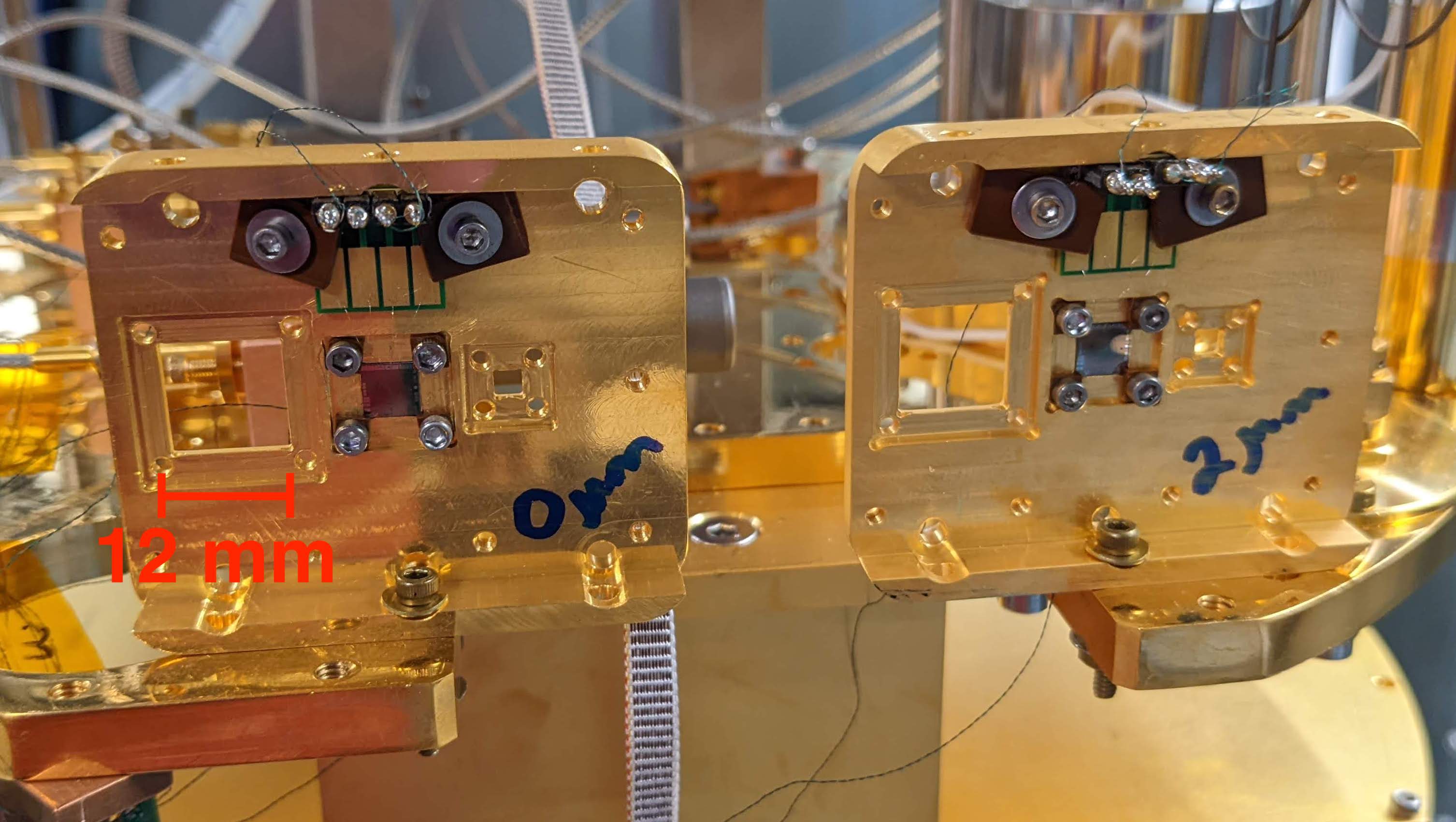}}
\caption{(a) A photograph of a 6~mm x 6~mm chip mounted in a gold-plated copper mounting jig for cosmic ray testing.  The jig and chip are precisely diced and machined such that the chip has a low thermal conductance to the jig as shown by the light making it through the 100$\mu$m gaps on each side of the chip. This helps maximize the cosmic ray events seen in the \glspl{tes} to increase the number of hits measured in a cryogenic run. (b) Two chips are installed side by side with identical mounting in the \gls{dr} to run comparison tests between two different mitigation configurations. The chip on the LHS has no Pd on the back side while the chip on the RHS is coated with two $\mu$m of Pd. \label{fig:testBed}}
\end{figure}

Simulation work from \cite{Stever:2021_JCAP} suggest that a layer of normal metal deposited on the backside of the device wafer should be effective at reducing the amplitude and rate of cosmic ray signals seen in bolometers.   The primary effect comes from the heat capacity and thermal conductivity of the Pd which helps keep the temperature of the substrate stable with minimal thermal gradients that could be caused by incident cosmic rays. 

We plan to measure the efficacy of cosmic ray mitigation as a function of Pd thickness on the back side of the chip.  Chips and wafers have been manufactured with 0, 0.5, and 2.0um of Pd and our currently running comparison tests at \gls{ucb} on both full detector wafers and individual test chips while methods of depositing thicker layers of Pd on the back side of the wafer continue to be explored. 

\section{Summary}

In these proceedings, we report on the design and fabrication status of the \glspl{lffpm} for the \lb\ satellite mission.  We describe the coupling architecture of the \glspl{fpm} in detail and the methodology we used to achieve it.  We are extending lenslet technology to be suitable for space flight by replacing traditional epoxy based \glspl{arc} with \gls{mars} laser-ablated into the surface of a silicon hemisphere.   We have made significant progress on the the LF-4 pixel type including the fabrication of bolometers that meet the specifications of \lb. 

\section{Acknowledgments}

The work in this proceeding is primarily funded by by NASA grant no. 80NSSC18K0132.  This work is also supported in Japan by ISAS/JAXA for Pre-Phase A2 studies, by the acceleration program of JAXA research and development directorate, by the World Premier International Research Center Initiative (WPI) of MEXT, by the JSPS Core-to-Core Program of A. Advanced Research Networks, and by JSPS KAKENHI Grant Numbers JP15H05891, JP17H01115, and JP17H01125. The Italian LiteBIRD phase A contribution is supported by the Italian Space Agency (ASI Grants No. 2020-9-HH.0 and 2016-24-H.1-2018), the National Institute for Nuclear Physics (INFN) and the National Institute for Astrophysics (INAF). The French LiteBIRD phase A contribution is supported by the Centre National d’Etudes Spatiale (CNES), by the Centre National de la Recherche Scientifique (CNRS), and by the Commissariat à l’Energie Atomique (CEA). The Canadian contribution is supported by the Canadian Space Agency.

Norwegian participation in LiteBIRD is supported by the Research Council of Norway (Grant No. 263011). The Spanish LiteBIRD phase A contribution is supported by the Spanish Agencia Estatal de Investigación (AEI), project refs. PID2019-110610RB-C21 and AYA2017-84185-P. Funds that support contributions from Sweden come from the Swedish National Space Agency (SNSA/Rymdstyrelsen) and the Swedish Research Council (Reg. no. 2019-03959). The German participation in LiteBIRD is supported in part by the Excellence Cluster ORIGINS, which is funded by the Deutsche Forschungsgemeinschaft (DFG, German Research Foundation) under Germany’s Excellence Strategy (Grant No. EXC-2094 - 390783311). This research used resources of the Central Computing System owned and operated by the Computing Research Center at KEK, as well as resources of the National Energy Research Scientific Computing Center, a DOE Office of Science User Facility supported by the Office of Science of the U.S. Department of Energy.

\textit{LiteBIRD} (phase A) activities are supported by the following funding sources: ISAS/JAXA, MEXT, JSPS, KEK (Japan); CSA (Canada); CNES, CNRS, CEA (France); DFG (Germany); ASI, INFN, INAF (Italy); RCN (Norway); AEI (Spain); SNSA, SRC (Sweden); NASA, DOE (USA).

\bibliography{Master_LiteBIRD_Detectors_SPIE_2022} 

\begin{thebibliography}{10}

\bibitem{LiteBIRD:2022_PTEP}
{LiteBIRD Collaboration}, E.~{Allys}, K.~{Arnold}, J.~{Aumont}, R.~{Aurlien},
  S.~{Azzoni}, C.~{Baccigalupi}, A.~J. {Banday}, R.~{Banerji}, R.~B.
  {Barreiro}, N.~{Bartolo}, L.~{Bautista}, D.~{Beck}, S.~{Beckman},
  M.~{Bersanelli}, F.~{Boulanger}, M.~{Brilenkov}, M.~{Bucher}, E.~{Calabrese},
  P.~{Campeti}, A.~{Carones}, F.~J. {Casas}, A.~{Catalano}, V.~{Chan},
  K.~{Cheung}, Y.~{Chinone}, S.~E. {Clark}, F.~{Columbro}, G.~{D'Alessandro},
  P.~{de Bernardis}, T.~{de Haan}, E.~{de la Hoz}, M.~{De Petris}, S.~{Della
  Torre}, P.~{Diego-Palazuelos}, T.~{Dotani}, J.~M. {Duval}, T.~{Elleflot},
  H.~K. {Eriksen}, J.~{Errard}, T.~{Essinger-Hileman}, F.~{Finelli},
  R.~{Flauger}, C.~{Franceschet}, U.~{Fuskeland}, M.~{Galloway}, K.~{Ganga},
  M.~{Gerbino}, M.~{Gervasi}, R.~T. {G{\'e}nova-Santos}, T.~{Ghigna},
  S.~{Giardiello}, E.~{Gjerl{\o}w}, J.~{Grain}, F.~{Grupp}, A.~{Gruppuso},
  J.~E. {Gudmundsson}, N.~W. {Halverson}, P.~{Hargrave}, T.~{Hasebe},
  M.~{Hasegawa}, M.~{Hazumi}, S.~{Henrot-Versill{\'e}}, B.~{Hensley}, L.~T.
  {Hergt}, D.~{Herman}, E.~{Hivon}, R.~A. {Hlozek}, A.~L. {Hornsby},
  Y.~{Hoshino}, J.~{Hubmayr}, K.~{Ichiki}, T.~{Iida}, H.~{Imada}, H.~{Ishino},
  G.~{Jaehnig}, N.~{Katayama}, A.~{Kato}, R.~{Keskitalo}, T.~{Kisner},
  Y.~{Kobayashi}, A.~{Kogut}, K.~{Kohri}, E.~{Komatsu}, K.~{Komatsu},
  K.~{Konishi}, N.~{Krachmalnicoff}, C.~L. {Kuo}, L.~{Lamagna}, M.~{Lattanzi},
  A.~T. {Lee}, C.~{Leloup}, F.~{Levrier}, E.~{Linder}, G.~{Luzzi},
  J.~{Macias-Perez}, B.~{Maffei}, D.~{Maino}, S.~{Mandelli},
  E.~{Mart{\'\i}nez-Gonz{\'a}lez}, S.~{Masi}, M.~{Massa}, S.~{Matarrese}, F.~T.
  {Matsuda}, T.~{Matsumura}, L.~{Mele}, M.~{Migliaccio}, Y.~{Minami},
  A.~{Moggi}, J.~{Montgomery}, L.~{Montier}, G.~{Morgante}, B.~{Mot},
  Y.~{Nagano}, T.~{Nagasaki}, R.~{Nagata}, R.~{Nakano}, T.~{Namikawa},
  F.~{Nati}, P.~{Natoli}, S.~{Nerval}, F.~{Noviello}, K.~{Odagiri}, S.~{Oguri},
  H.~{Ohsaki}, L.~{Pagano}, A.~{Paiella}, D.~{Paoletti}, A.~{Passerini},
  G.~{Patanchon}, F.~{Piacentini}, M.~{Piat}, G.~{Polenta}, D.~{Poletti},
  T.~{Prouv{\'e}}, G.~{Puglisi}, D.~{Rambaud}, C.~{Raum}, S.~{Realini},
  M.~{Reinecke}, M.~{Remazeilles}, A.~{Ritacco}, G.~{Roudil}, J.~A.
  {Rubino-Martin}, M.~{Russell}, H.~{Sakurai}, Y.~{Sakurai}, M.~{Sasaki},
  D.~{Scott}, Y.~{Sekimoto}, K.~{Shinozaki}, M.~{Shiraishi}, P.~{Shirron},
  G.~{Signorelli}, F.~{Spinella}, S.~{Stever}, R.~{Stompor}, S.~{Sugiyama},
  R.~M. {Sullivan}, A.~{Suzuki}, T.~L. {Svalheim}, E.~{Switzer}, R.~{Takaku},
  H.~{Takakura}, Y.~{Takase}, A.~{Tartari}, Y.~{Terao}, J.~{Thermeau},
  H.~{Thommesen}, K.~L. {Thompson}, M.~{Tomasi}, M.~{Tominaga}, M.~{Tristram},
  M.~{Tsuji}, M.~{Tsujimoto}, L.~{Vacher}, P.~{Vielva}, N.~{Vittorio},
  W.~{Wang}, K.~{Watanuki}, I.~K. {Wehus}, J.~{Weller}, B.~{Westbrook},
  J.~{Wilms}, E.~J. {Wollack}, J.~{Yumoto}, and M.~{Zannoni}, ``{Probing Cosmic
  Inflation with the LiteBIRD Cosmic Microwave Background Polarization
  Survey},'' {\em arXiv e-prints} , p.~arXiv:2202.02773, Feb. 2022.

\bibitem{LiteBIRD:2021_Hazumi}
M.~{Hazumi}, P.~A.~R. {Ade}, A.~{Adler}, E.~{Allys}, K.~{Arnold}, D.~{Auguste},
  J.~{Aumont}, R.~{Aurlien}, J.~{Austermann}, C.~{Baccigalupi}, A.~J. {Banday},
  R.~{Banjeri}, R.~B. {Barreiro}, S.~{Basak}, J.~{Beall}, D.~{Beck},
  S.~{Beckman}, J.~{Bermejo}, P.~{de Bernardis}, M.~{Bersanelli}, J.~{Bonis},
  J.~{Borrill}, F.~{Boulanger}, S.~{Bounissou}, M.~{Brilenkov}, M.~{Brown},
  M.~{Bucher}, E.~{Calabrese}, P.~{Campeti}, A.~{Carones}, F.~J. {Casas},
  A.~{Challinor}, V.~{Chan}, K.~{Cheung}, Y.~{Chinone}, J.~F. {Cliche},
  L.~{Colombo}, F.~{Columbro}, J.~{Cubas}, A.~{Cukierman}, D.~{Curtis},
  G.~{D'Alessandro}, N.~{Dachlythra}, M.~{De Petris}, C.~{Dickinson},
  P.~{Diego-Palazuelos}, M.~{Dobbs}, T.~{Dotani}, L.~{Duband}, S.~{Duff}, J.~M.
  {Duval}, K.~{Ebisawa}, T.~{Elleflot}, H.~K. {Eriksen}, J.~{Errard},
  T.~{Essinger-Hileman}, F.~{Finelli}, R.~{Flauger}, C.~{Franceschet},
  U.~{Fuskeland}, M.~{Galloway}, K.~{Ganga}, J.~R. {Gao}, R.~{Genova-Santos},
  M.~{Gerbino}, M.~{Gervasi}, T.~{Ghigna}, E.~{Gjerl{\o}w}, M.~L. {Gradziel},
  J.~{Grain}, F.~{Grupp}, A.~{Gruppuso}, J.~E. {Gudmundsson}, T.~{de Haan},
  N.~W. {Halverson}, P.~{Hargrave}, T.~{Hasebe}, M.~{Hasegawa}, M.~{Hattori},
  S.~{Henrot-Versill{\'e}}, D.~{Herman}, D.~{Herranz}, C.~A. {Hill},
  G.~{Hilton}, Y.~{Hirota}, E.~{Hivon}, R.~A. {Hlozek}, Y.~{Hoshino}, E.~{de la
  Hoz}, J.~{Hubmayr}, K.~{Ichiki}, T.~{Iida}, H.~{Imada}, K.~{Ishimura},
  H.~{Ishino}, G.~{Jaehnig}, T.~{Kaga}, S.~{Kashima}, N.~{Katayama}, A.~{Kato},
  T.~{Kawasaki}, R.~{Keskitalo}, T.~{Kisner}, Y.~{Kobayashi}, N.~{Kogiso},
  A.~{Kogut}, K.~{Kohri}, E.~{Komatsu}, K.~{Komatsu}, K.~{Konishi},
  N.~{Krachmalnicoff}, I.~{Kreykenbohm}, C.~L. {Kuo}, A.~{Kushino},
  L.~{Lamagna}, J.~V. {Lanen}, M.~{Lattanzi}, A.~T. {Lee}, C.~{Leloup},
  F.~{Levrier}, E.~{Linder}, T.~{Louis}, G.~{Luzzi}, T.~{Maciaszek},
  B.~{Maffei}, D.~{Maino}, M.~{Maki}, S.~{Mandelli}, E.~{Martinez-Gonzalez},
  S.~{Masi}, T.~{Matsumura}, A.~{Mennella}, M.~{Migliaccio}, Y.~{Minami},
  K.~{Mitsuda}, J.~{Montgomery}, L.~{Montier}, G.~{Morgante}, B.~{Mot},
  Y.~{Murata}, J.~A. {Murphy}, M.~{Nagai}, Y.~{Nagano}, T.~{Nagasaki},
  R.~{Nagata}, S.~{Nakamura}, T.~{Namikawa}, P.~{Natoli}, S.~{Nerval},
  T.~{Nishibori}, H.~{Nishino}, F.~{Noviello}, C.~{O'Sullivan}, H.~{Ogawa},
  H.~{Ogawa}, S.~{Oguri}, H.~{Ohsaki}, I.~S. {Ohta}, N.~{Okada}, N.~{Okada},
  L.~{Pagano}, A.~{Paiella}, D.~{Paoletti}, G.~{Patanchon}, J.~{Peloton},
  F.~{Piacentini}, G.~{Pisano}, G.~{Polenta}, D.~{Poletti}, T.~{Prouv{\'e}},
  G.~{Puglisi}, D.~{Rambaud}, C.~{Raum}, S.~{Realini}, M.~{Reinecke},
  M.~{Remazeilles}, A.~{Ritacco}, G.~{Roudil}, J.~A. {Rubino-Martin},
  M.~{Russell}, H.~{Sakurai}, Y.~{Sakurai}, M.~{Sandri}, M.~{Sasaki},
  G.~{Savini}, D.~{Scott}, J.~{Seibert}, Y.~{Sekimoto}, B.~{Sherwin},
  K.~{Shinozaki}, M.~{Shiraishi}, P.~{Shirron}, G.~{Signorelli}, G.~{Smecher},
  S.~{Stever}, R.~{Stompor}, H.~{Sugai}, S.~{Sugiyama}, A.~{Suzuki},
  J.~{Suzuki}, T.~L. {Svalheim}, E.~{Switzer}, R.~{Takaku}, H.~{Takakura},
  S.~{Takakura}, Y.~{Takase}, Y.~{Takeda}, A.~{Tartari}, E.~{Taylor},
  Y.~{Terao}, H.~{Thommesen}, K.~L. {Thompson}, B.~{Thorne}, T.~{Toda},
  M.~{Tomasi}, M.~{Tominaga}, N.~{Trappe}, M.~{Tristram}, M.~{Tsuji},
  M.~{Tsujimoto}, C.~{Tucker}, J.~{Ullom}, G.~{Vermeulen}, P.~{Vielva},
  F.~{Villa}, M.~{Vissers}, N.~{Vittorio}, I.~{Wehus}, J.~{Weller},
  B.~{Westbrook}, J.~{Wilms}, B.~{Winter}, E.~J. {Wollack}, N.~Y. {Yamasaki},
  T.~{Yoshida}, J.~{Yumoto}, M.~{Zannoni}, and A.~{Zonca}, ``{LiteBIRD
  satellite: JAXA's new strategic L-class mission for all-sky surveys of cosmic
  microwave background polarization},'' in {\em Society of Photo-Optical
  Instrumentation Engineers (SPIE) Conference Series},  {\em Society of
  Photo-Optical Instrumentation Engineers (SPIE) Conference Series} {\bf
  11443}, p.~114432F, Dec. 2020.

\bibitem{Jaehnig:2022_SPIE}
G.~{Jaehnig} {\em Submitted to SPIE 2022} , 2022.

\bibitem{Bleem:2015_SPTSZGalaxies}
L.~E. Bleem {\em et~al.}, ``{Galaxy Clusters Discovered via the
  Sunyaev-Zel'dovich Effect in the 2500-square-degree SPT-SZ survey},'' {\em
  Astrophys. J. Suppl.}~{\bf 216}(2), p.~27, 2015.

\bibitem{Abitbol:2018_EBEX}
M.~Abitbol {\em et~al.}, ``The {EBEX} balloon-borne
  experiment{\textemdash}detectors and readout,'' {\em Astrophys. J.
  Suppl.}~{\bf 239}, p.~8, Nov. 2018.

\bibitem{Arnold:2010_POLARBEAR}
K.~Arnold, P.~A. Ade, A.~Anthony, F.~Aubin, D.~Boettger, J.~Borrill,
  C.~Cantalupo, M.~Dobbs, J.~Errard, D.~Flanigan, {\em et~al.}, ``The polarbear
  cmb polarization experiment,'' in {\em Millimeter, Submillimeter, and
  Far-Infrared Detectors and Instrumentation for Astronomy V},   {\bf 7741},
  p.~77411E, International Society for Optics and Photonics, 2010.

\bibitem{Suzuki:2016_PB2SA}
A.~{Suzuki}, P.~{Ade}, Y.~{Akiba}, C.~{Aleman}, K.~{Arnold}, C.~{Baccigalupi},
  B.~{Barch}, D.~{Barron}, A.~{Bender}, D.~{Boettger}, J.~{Borrill},
  S.~{Chapman}, Y.~{Chinone}, A.~{Cukierman}, M.~{Dobbs}, A.~{Ducout},
  R.~{Dunner}, T.~{Elleflot}, J.~{Errard}, G.~{Fabbian}, S.~{Feeney},
  C.~{Feng}, T.~{Fujino}, G.~{Fuller}, A.~{Gilbert}, N.~{Goeckner-Wald},
  J.~{Groh}, T.~D. {Haan}, G.~{Hall}, N.~{Halverson}, T.~{Hamada},
  M.~{Hasegawa}, K.~{Hattori}, M.~{Hazumi}, C.~{Hill}, W.~{Holzapfel},
  Y.~{Hori}, L.~{Howe}, Y.~{Inoue}, F.~{Irie}, G.~{Jaehnig}, A.~{Jaffe},
  O.~{Jeong}, N.~{Katayama}, J.~{Kaufman}, K.~{Kazemzadeh}, B.~{Keating},
  Z.~{Kermish}, R.~{Keskitalo}, T.~{Kisner}, A.~{Kusaka}, M.~L. {Jeune},
  A.~{Lee}, D.~{Leon}, E.~{Linder}, L.~{Lowry}, F.~{Matsuda}, T.~{Matsumura},
  N.~{Miller}, K.~{Mizukami}, J.~{Montgomery}, M.~{Navaroli}, H.~{Nishino},
  J.~{Peloton}, D.~{Poletti}, G.~{Puglisi}, G.~{Rebeiz}, C.~{Raum},
  C.~{Reichardt}, P.~{Richards}, C.~{Ross}, K.~{Rotermund}, Y.~{Segawa},
  B.~{Sherwin}, I.~{Shirley}, P.~{Siritanasak}, N.~{Stebor}, R.~{Stompor},
  J.~{Suzuki}, O.~{Tajima}, S.~{Takada}, S.~{Takakura}, S.~{Takatori},
  A.~{Tikhomirov}, T.~{Tomaru}, B.~{Westbrook}, N.~{Whitehorn}, T.~{Yamashita},
  A.~{Zahn}, and O.~{Zahn}, ``{The Polarbear-2 and the Simons Array
  Experiments},'' {\em Journal of Low Temperature Physics}~{\bf 184},
  pp.~805--810, Aug. 2016.

\bibitem{Abitbol:2019_SODecadal}
M.~H. Abitbol {\em et~al.}, ``{The Simons Observatory: Astro2020 Decadal
  Project Whitepaper},'' {\em Bull. Am. Astron. Soc.}~{\bf 51}, p.~147, 2019.

\bibitem{Ghinga:2022_SPIE}
T.~{Ghigna} {\em Submitted to SPIE 2022} , 2022.

\bibitem{Russell:2022_SPIE}
M.~{Russell} {\em Submitted to SPIE 2022} , 2022.

\bibitem{Westbrook:2018_SimonsArray}
B.~Westbrook, P.~A.~R. Ade, M.~Aguilar, Y.~Aklba, K.~Arnold, C.~Baccigalupi,
  D.~Barron, D.~Beck, S.~Beckman, and A.~N. Bender, ``The polarbear-2 and
  simons array focal plane fabrication status,'' {\em Journal of Low
  Temperature Physics} (5).

\bibitem{Carter:2018_SPT3G2year}
F.~W. Carter, T.~W. Cecil, C.~L. Chang, H.-M. Cho, J.-F. Cliche, T.~M.
  Crawford, A.~Cukierman, E.~V. Denison, T.~de~Haan, J.~Ding, M.~A. Dobbs,
  D.~Dutcher, W.~Everett, A.~Foster, J.~C. Groh, A.~Gilbert, N.~W. Halverson,
  A.~H. Harke-Hosemann, N.~L. Harrington, J.~W. Henning, G.~C. Hilton, G.~P.
  Holder, W.~L. Holzapfel, N.~Huang, K.~D. Irwin, O.~B. Jeong, M.~Jonas, T.~S.
  Khaire, L.~Knox, A.~M. Kofman, M.~Korman, D.~L. Kubik, S.~Kuhlmann, C.-L.
  Kuo, A.~T. Lee, E.~M. Leitch, A.~E. Lowitz, S.~S. Meyer, D.~Michalik,
  J.~Montgomery, A.~Nadolski, T.~Natoli, H.~Ngyuen, G.~I. Noble, V.~Novosad,
  S.~Padin, Z.~Pan, J.~Pearson, C.~M. Posada, A.~Rahlin, C.~L. Reichardt, J.~E.
  Ruhl, J.~T. Sayre, E.~Shirokoff, G.~Smecher, J.~A. Sobrin, A.~A. Stark, K.~T.
  Story, A.~Suzuki, K.~L. Thompson, C.~Tucker, L.~R. Vale, K.~Vanderlinde,
  J.~Vieira, G.~Wang, N.~Whitehorn, V.~Yefremenko, K.~W. Yoon, M.~Young, A.~N.
  Bender, P.~A.~R. Ade, Z.~Ahmed, A.~J. Anderson, J.~S. Avva, P.~S. Barry,
  R.~B. Thakur, B.~A. Benson, L.~S. Bleem, K.~Byrum, J.~E. Carlstrom, K.~Aylor,
  S.~Bocquet, S.~Dodelson, J.~Gallicchio, S.~Guns, W.~Quan, S.~Raghunathan,
  W.~L. Wu, and A.~Jones, ``Year two instrument status of the {SPT}-3g cosmic
  microwave background receiver,'' in {\em Millimeter, Submillimeter, and
  Far-Infrared Detectors and Instrumentation for Astronomy {IX}},
  J.~Zmuidzinas and J.-R. Gao, eds., {SPIE}, Aug. 2018.

\bibitem{Suzuki:2013_Thesis}
A.~Suzuki, ``{Multichroic Bolometric Detector Architecture for Cosmic Microwave
  Background Polarimetry Experiments},'' {\em PhD thesis, University of
  California, Berkeley} , 2013.

\bibitem{Edwards:2012_LensletSinuous}
J.~M. Edwards, R.~O'Brient, A.~T. Lee, and G.~M. Rebeiz, ``Dual-polarized
  sinuous antennas on extended hemispherical silicon lenses,'' {\em IEEE
  Transactions on Antennas and Propagation}~{\bf 60}(9), pp.~4082--4091, 2012.

\bibitem{Farias:2022_MARS}
N.~Farias, S.~Beckman, A.~T. Lee, and A.~Suzuki, ``Simulated performance of
  laser-machined metamaterial anti-reflection coatings,'' {\em Journal of Low
  Temperature Physics} , 2022.

\bibitem{Westbrook:2020_SPIE}
B.~Westbrook, C.~Raum, S.~Beckman, A.~T. Lee, N.~Farias, T.~Sasse, A.~Suzuki,
  E.~Kane, J.~E. Austermann, J.~A. Beall, S.~M. Duff, J.~Hubmayr, G.~C. Hilton,
  J.~V. Lanen, M.~R. Vissers, M.~R. Link, N.~Halverson, G.~Jaehnig, T.~Ghinga,
  S.~Stever, Y.~Minami, K.~L. Thompson, M.~Russell, K.~Arnold, J.~Seibert, and
  M.~Silva-Feaver, ``{Detector fabrication development for the LiteBIRD
  satellite mission},'' in {\em Space Telescopes and Instrumentation 2020:
  Optical, Infrared, and Millimeter Wave},  M.~Lystrup, M.~D. Perrin,
  N.~Batalha, N.~Siegler, and E.~C. Tong, eds.,  {\bf 11443}, pp.~915 -- 936,
  International Society for Optics and Photonics, SPIE, 2020.

\bibitem{Miniussi:2014_Planck_CosmicRays}
A.~{Miniussi}, J.~L. {Puget}, W.~{Holmes}, G.~{Patanchon}, A.~{Catalano},
  Y.~{Giraud-Heraud}, F.~{Pajot}, M.~{Piat}, and L.~{Vibert}, ``{Study of
  Cosmic Ray Impact on Planck/HFI Low Temperature Detectors},'' {\em Journal of
  Low Temperature Physics}~{\bf 176}, pp.~815--821, Sept. 2014.

\bibitem{Stever:2021_JCAP}
S.~L. {Stever}, T.~{Ghigna}, M.~{Tominaga}, G.~{Puglisi}, M.~{Tsujimoto},
  M.~{Zeccoli Marazzini}, M.~{Baratto}, M.~{Tomasi}, Y.~{Minami},
  S.~{Sugiyama}, A.~{Kato}, T.~{Matsumura}, H.~{Ishino}, G.~{Patanchon}, and
  M.~{Hazumi}, ``{Simulations of systematic effects arising from cosmic rays in
  the LiteBIRD space telescope, and effects on the measurements of CMB
  B-modes},'' {\em JCAP}~{\bf 2021}, p.~013, Sept. 2021.

\end{thebibliography}

\bibliographystyle{spiebib} 

\end{document}